\documentclass[%
 reprint,
superscriptaddress,
floatfix,
 amsmath,amssymb,
 aps, prx, noeprint
]{revtex4-2}

\usepackage[utf8]{inputenc}
\usepackage{amsmath,esint}
\usepackage{mathtools}
\usepackage[colorlinks=true,linkcolor=blue,citecolor=blue,urlcolor=blue]{hyperref}
\usepackage{amsfonts}
\usepackage{amssymb}
\usepackage{bm}
\usepackage{textcomp}
\usepackage{empheq}
\usepackage{siunitx}
\usepackage[titletoc,title]{appendix}
\usepackage{graphicx}
\usepackage{dcolumn}
\usepackage{bm}
\usepackage{subcaption}
\usepackage{xcolor,soul}
\usepackage{lipsum,afterpage,refcount}

\renewcommand{\Re}{\operatorname{Re}}
\renewcommand{\Im}{\operatorname{Im}}

\newcommand{\citeasnoun}[1]{Ref.~\cite{#1}}

\newcommand{\Figref}[1]{Figure~\ref{fig:#1}}
\newcommand{\figref}[1]{Fig.~\ref{fig:#1}}
\renewcommand{\eqref}[1]{Eq.~(\ref{eq:#1})}
\newcommand{\Eqref}[1]{Equation~(\ref{eq:#1})}
\newcommand{\eqreftwo}[2]{Eqs.~(\ref{eq:#1},\ref{eq:#2})}

\newcommand{\secref}[1]{Sec.~\ref{sec:#1}}

\newcommand*{\Ev}{\mathbf{E}}
\newcommand*{\Hv}{\mathbf{H}}
\newcommand*{\Dv}{\mathbf{D}}
\newcommand*{\Bv}{\mathbf{B}}

\newcommand*{\kv}{\mathbf{k}}

\newcommand*{\SM}{SM}

\newcommand*{\epsnl}{\overline{\overline{\varepsilon_{\textrm{nl}}}}(\omega,\mathbf{k})}
\newcommand*{\chinl}{\overline{\overline{\chi_{\textrm{nl}}}}(\omega,\mathbf{k})}
\newcommand*{\epsten}{\overline{\overline{\varepsilon}}}
\newcommand*{\muten}{\overline{\overline{\mu}}}
\newcommand*{\xiten}{\overline{\overline{\xi}}}
\newcommand*{\zetaten}{\overline{\overline{\zeta}}}
\newcommand*{\id}{\overline{\overline{\mathbf{I}}}}
\newcommand*{\ts}{\textsubscript}

\begin{document}

\preprint{APS/123-QED}

\title{Fundamental limits to the refractive index of transparent optical materials}

\author{Hyungki Shim}
\affiliation{Department of Applied Physics and Energy Sciences Institute, Yale University, New Haven, Connecticut 06511, USA}
\affiliation{Department of Physics, Yale University, New Haven, Connecticut 06511, USA}
\author{Francesco Monticone}
\affiliation{School of Electrical and Computer Engineering, Cornell University, Ithaca, New York 14853, USA}
\author{Owen D. Miller}
\email{owen.miller@yale.edu}
\affiliation{Department of Applied Physics and Energy Sciences Institute, Yale University, New Haven, Connecticut 06511, USA}

\date{\today}

\begin{abstract}
    Increasing the refractive index available for optical and nanophotonic systems opens new vistas for design: for applications ranging from broadband metalenses to ultrathin photovoltaics to high-quality-factor resonators, higher index directly leads to better devices with greater functionality. Although standard transparent materials have been limited to refractive indices smaller than 3 in the visible, recent metamaterials designs have achieved refractive indices above 5, accompanied by high losses, and near the phase transition of a ferroelectric perovskite a broadband index above 26 has been claimed. In this work, we derive fundamental limits to the refractive index of any material, given only the underlying electron density and either the maximum allowable dispersion or the minimum bandwidth of interest. The Kramers--Kronig relations provide a representation for any passive (and thereby causal) material, and a well-known sum rule constrains the possible distribution of oscillator strengths. In the realm of small to modest dispersion, our bounds are closely approached and not surpassed by a wide range of natural materials, showing that nature has already nearly reached a Pareto frontier for refractive index and dispersion. Surprisingly, our bound shows a cube-root dependence on electron density, meaning that a refractive index of 26 over all visible frequencies is likely impossible. Conversely, for narrow-bandwidth applications, nature does not provide the highly dispersive, high-index materials that our bounds suggest should be possible. We use the theory of composites to identify metal-based metamaterials that can exhibit small losses and sizeable increases in refractive index over the current best materials. Moreover, if the ``elusive lossless metal'' can be synthesized, we show that it would enable arbitrarily high refractive index in the high-dispersion regime, nearly achieving our bounds even at refractive indices of 100 and beyond at optical frequencies.
\end{abstract}

\pacs{Valid PACS appear here}
\maketitle

Increasing the refractive index of optical materials would unlock new levels of functionality in fields ranging from metasurface optics~\cite{Zheludev2012a,Yu2014a,Lin2014,Arbabi2015,Capasso2016,Kuznetsov2016a,Li2018a,Presutti2020} to high-quality-factor resonators~\cite{Noda2000c,Michler2000,Loncar2003,Reithmaier2004b,Tanabe2005,Hennessy2007b,Srinivasan2007b}. In this Article, we develop a framework for identifying fundamental limits to the maximum possible refractive index in any material or metamaterial, dependent only on the achievable electron density, the frequency range of interest, and possibly a maximum allowable dispersion. We show that the Kramers--Kronig relations for optical susceptibilities, in conjunction with a well-known sum rule, impose surprisingly strong constraints on refractive-index lineshapes, imposing strict limitations to refractive index at high frequency, with only weak (cube-root) increases possible through electron-density enhancements or large allowable dispersion. We show that a large range of questions around maximum index, including bandwidth-averaged objectives with constraints on dispersion and/or loss, over the entire range of causality-allowed refractive indices, can be formulated as linear programs amenable to computational global bounds, and that many questions of interest have global bounds with optima that are single Drude--Lorentz oscillators, leading to simple analytical bounds. For the central question of maximum index at any given frequency, we show that many natural materials already closely approach the Pareto frontier of tradeoffs with density, dispersion, and frequency, with little room (ranging from 1.1--1.5$\times$) for significant improvement. 
We apply our framework to high-index optical glasses (characterized by their Abbe number) and bandwidth-based bounds. For anisotropic refractive indices, or materials with magnetic in addition to electric response, we use a nonlocal-medium-based transformation to prove that any positive- or negative-semidefinite material properties cannot surpass these bounds, although there is an intriguing loophole for hyperbolic metamaterials. At optical frequencies, there are few or no natural materials with high index and high dispersion, but we show that composite metamaterials can be designed to have refractive indices approaching our bounds. With conventional metals such as gold and aluminum, we show that low-loss refractive indices of 5 in the visible, 18 in the near-infrared (\SI{3}{\um} wavelength), and 40 in the mid-infrared (\SI{10}{\um} wavelength) are achievable. If a near-zero-loss metal can be discovered or synthesized~\cite{Khurgin2010,Boltasseva2011}, high-dispersion refractive indices above 100 would be possible at any optical frequency. 

A large material refractive index $n$ offers significant benefits for nanophotonics devices. First, the reduced internal wavelength enables rapid phase oscillations, which enable wavefront reshaping over short distances and is the critical requirement of high-efficiency metalenses and metasurfaces~\cite{Zheludev2012a,Yu2014a,Lin2014,Arbabi2015,Capasso2016,Kuznetsov2016a,Li2018a,Presutti2020}. Second, it dramatically increases the internal photon density of states, which scales as $n^3$ in a bulk material~\cite{Yariv1989} and offers the possibility for greater tunability and functionality. The enhanced density of states is responsible for the ray-optical $4n^2$ ``Yablonovitch limit'' to all-angle solar absorption~\cite{Yablonovitch1982a} and the random surface textures employed in commercial photovoltaics. Third, high optical index unlocks the capability for near-degenerate electric and magnetic resonances within nano-resonators. Tandem electric and magnetic response is critical for highly directional control of waves; whereas a single electric dipole radiates equally into opposite directions, a tandem electric and magnetic dipole can radiate efficiently into a single, controllable direction, known as the ``Kerker effect,"~\cite{Kerker1983} then forming the building blocks of complex, tailored scattering profiles~\cite{Schuller2007,Ginn2012,Geffrin2012,Fu2013,Person2013,Bakker2015,Monticone2017,Vaskin2018,Komar2018}. Bound states in the continuum utilize Kerker-like phenomena and may also benefit from high index~\cite{Doeleman2018}. Fourth, a large phase index can lead to a large group index, which underpins the entire field of slow light~\cite{Baba2008,Krauss2008}, for applications from delay lines to compressing optical signals. Finally, high refractive index enables significant reductions of the smallest possible mode volume in a dielectric resonator. Recent theoretical and experimental demonstrations show the possibility for highly subwavelength mode volumes in lossless dielectric materials~\cite{Robinson2005,Liang2013,Hu2016a,Choi2017,Zhao2020}. In this case, a high refractive index increases the discontinuities in the electric and displacement fields across small-feature boundaries, enabling significant enhancements of the local field intensity that are useful for applications from single-molecule imaging~\cite{Nie1997a, Kneipp1997, VanZanten2009, Schermelleh2010} to high-efficiency nonlinear frequency conversion~\cite{Marcy1995,Jain1996,Merriam2000,Petrov2010}. 

The very highest refractive indices of transparent natural materials are 4 to 4.2 at near-infrared frequencies~\cite{Li1980}, and 2.85 at visible frequencies~\cite{Devore1951}. Metamaterials, comprising multiple materials combined in random or designed patterns, have been designed with refractive indices up to 5 at visible frequencies~\cite{Kim2016}, albeit with significant material losses. As the frequency is reduced, the refractive index can be significantly increased, a feature predicted by our bounds and borne out by the literature. Low-loss metamaterials have been designed to achieve refractive indices near 7 at infrared frequencies (3--\SI{6}{\um} wavelengths)~\cite{Shin2009} and above 38 at terahertz frequencies~\cite{Choi2011}. Near the phase transition of ferroelectric materials, it is known (Chap. 16 of \citeasnoun{Kittel1996}) that in principle the refractive index is unlimited. Yet the caveat is that the frequency at which this occurs must go to zero. Experimental and theoretical studies have identified multiple materials with ``colossal'' zero-frequency (electrostatic) dielectric constants~\cite{Lunkenheimer2009}, even surpassing values of 10,000~\cite{Hu2013}. All of these results are consistent with and predicted by the bounds that we derive.

Recently, scattering measurements on the perovskite material KTN:Li near its phase transition led to the claim of a refractive index of at least 26 across the entire visible region~\cite{DiMei2018}. As we discuss further below, such a refractive index appears to be theoretically impossible: it would require an electron density and/or dispersion almost three orders of magnitude larger than those of known materials, an unprecedented anomaly. Thus our work suggests that the experimental measurements may arise from linear-diffraction or even nonlinear optical effects, and do not represent a true phase-delay refractive index.

Theoretical inquiries into possible refractive indices have revolved around \emph{models} that relate refractive index to other material properties, and particularly that of the energy gap in a semiconducting or insulating material. The well-known Moss Relation~\cite{Moss1950,Ravindra2007} is a heuristic model that suggests that refractive index falls off as the fourth root of the energy gap of the material. This model can effectively describe some materials over a limited energy range, but is not a rigorous relation and cannot be used for definite bounds. Another approach, related to ours, is to use the Kramers--Kronig relation for refractive index to suggest that refractive index should scale with the square root of electron density~\cite{Jackson1999,Andreoli2021}. But this approach has not been used for definite bounds, nor is the scaling relation correct: as we show, an alternative susceptibility-based sum rule shows that the refractive index should scale as the cube-root of electron density (for a fixed dispersion value, without which refractive index can in principle be arbitrarily high). A recent result utilizes renormalization-group theory to suggest that the refractive index of an ensemble of atoms must saturate around 1.7 (\citeasnoun{Andreoli2021}). There have also been bounds on \emph{nonlinear} susceptibilities using quantum-mechanical sum rules~\cite{Kuzyk2000,Kuzyk2006}, but, as far as we know, there have not been bounds for arbitrary materials on linear refractive index, which is the key controlling property for optics and nanophotonics applications.

Separately, bounds \emph{have} been developed for other material properties, such as the minimum dispersion of a negative-permittivity or negative-index material~\cite{Skaar2006,Gustafsson2010}. Such bounds utilize causality properties, similar to our work, to optimize over all possible susceptibility functions. There have also been claims of bounds on the minimum losses of a negative-refraction material~\cite{Stockman2007}, though recent work~\cite{Milton2020} has identified errors in that reasoning and shown that lossless negative-refraction materials are possible, in principle. If the approaches of these papers were directly applied to refractive index, they would yield trivially infinite bounds, as they do not make use of the electron-density sum rule of \eqref{fsum} below. A large range of electromagnetic response functions have recently been bounded through analytical or computational approaches~\cite{miller_polimeridis_reid_hsu_delacy_joannopoulos_soljacic_johnson_2016,hugonin_besbes_ben-abdallah_2015, miller_ilic_christensen_reid_atwater_joannopoulos_soljacic_johnson_2017, Shim2019, Zhang2019, Angeris2019, Molesky2019a, Gustafsson2019,Molesky2020a, Trivedi2020, Molesky2020b, Kuang2020, Kuang2020b}, but none of these approaches have been applied to refractive index, nor is there a clear pathway to do so.

In this paper, we establish the maximal attainable refractive index for arbitrary passive, linear, bianisotropic media, applicable to naturally occurring materials as well as artificial metamaterials. We first derive a general representation of optical susceptibility starting from the Kramers--Kronig relations (\secref{linprog}), enabling us to describe any material by a sum of Drude--Lorentz oscillators with infinitesimal loss rates (\secref{linprog}). By considering a design space of an arbitrarily large number of oscillators, the susceptibility is a linear function of the degrees of freedom, which are the oscillator strengths. Many constraints (dispersion, bandwidth, loss rate, etc.) are also linear functions of the oscillator strengths, which themselves are constrained by the electron density via a well-known sum rule. Thus a large set of questions around maximum refractive index are \emph{linear programs}, whose global optima can be computed quickly and efficiently~\cite{Vanderbei2015}. The canonical question is: what is the largest possible refractive index at any frequency $\omega$, such that the material dispersion is bounded? In \secref{sf} we show that this linear program has an \emph{analytical} bound, which is a single, lossless Drude--Lorentz oscillator (corresponding to all oscillator strengths being concentrated at a single electronic transition). These bounds describe universal tradeoffs between refractive index, dispersion, and frequency, and we show that many natural materials and metamaterials closely approach the bounds. We then devote a separate section (\secref{abbe}) to optical glasses, which are highly studied and critical for high-quality optical components. We show that our bounds closely describe the behavior of such glasses, and that there may be opportunities for improvement at low Abbe numbers (high dispersion values). An alternative characterization for refractive index may not be a specific dispersion value, and instead a desired bandwidth of operation, and in \secref{bw} we derive bounds on refractive index as a function of allowable bandwidth. Across all of our bounds we find that there may be small improvements possible relative to current materials (1.1--1.5$\times$). Finally, we consider the possibilities of anisotropy, magnetic permeability, and/or magneto-electric coupling in \secref{mag}. We show that a large swath of such effects cannot lead to higher refractive indices, and are subject to the same isotropic-index bounds derived earlier in the paper. We also find intriguing loopholes including gyrotropic plasmonic media (which have a modified Kramers--Kronig relation) and hyperbolic metamaterials, although the former may be particularly hard to achieve at optical frequencies while the high-index modes in the latter may be difficult to access for free-space propagating plane waves. We identify exactly the material properties that enable such loopholes. Furthermore, we use the theory of composites to design low loss, highly dispersive, metal-based metamaterials with higher indices than have ever been measured or designed (\secref{highindex}). In the Conclusion, \secref{conclusion}, we discuss possible extensions of our framework to incorporate alternative metrics, gain media, anomalous dispersion, and nonlinear response.
	  
\section{Maximum refractive index as a linear program} \label{sec:linprog}
	 		 
To identify the maximal refractive index, one first needs a representation of all physically allowable material susceptibilities. We consider here a transparent, isotropic, nonmagnetic material, which can be described by its refractive index $n$, relative permittivity $\varepsilon = n^2$, or its susceptibility $\chi = \varepsilon - 1$. (We discuss extensions to anisotropic and/or magnetic materials in \secref{mag} and we discuss the possible inclusion of loss below.) Instead of assuming a particular form for the susceptibilities (like a small number of Drude--Lorentz oscillators), we assume only passivity: that the polarization currents in the material do no net work. Any passive material must be causal~\cite{Nussenzveig1972}; causality, alongside technical conditions on the appropriate behavior at infinitely large frequencies on the real axis, implies that each of the material parameters must satisfy the \emph{Kramers--Kronig} (KK) relations. One version of the KK relation for the material susceptibility relates its real part at one frequency to a principal-value integral of its imaginary part over all frequencies~\cite{Lucarini2005}:
\begin{align}
\Re \chi(\omega) = \frac{2}{\pi}  \int_0^\infty \frac{\omega' \Im \chi(\omega')}{\omega'^2 - \omega^2} \,{\rm d}\omega'.  \label{eq:kk} 
\end{align}
Any isotropic material's susceptibility must satisfy \eqref{kk}. The existence of KK relations, together with passivity restrictions, already imply bounds on minimum dispersion in regions of negative refractive index~\cite{Skaar2006,Stockman2007,Gustafsson2010}, but it does not by itself impose any bound on how large the real part of the susceptibility (and correspondingly the refractive index) can be. The key constraint is the ``$f$-sum rule:'' a certain integral of the imaginary part of the susceptibility \emph{must} equal a particular constant multiplied by the electron density $N_e$ of the medium. Typically, electron density is folded into a frequency $\omega_p$, which for metals is the plasma frequency but for any material describes the high-frequency asymptotic response of the material. The $f$-sum rule for the susceptibility is~\cite{Thomas1925,Kuhn1925,King1976,Lucarini2005}
\begin{align}
\int_0^\infty \omega' \Im \chi(\omega') \,{\rm d}\omega' = \frac{\pi e^2 N_e}{2\varepsilon_0 m_e} = \frac{\pi \omega_p^2}{2} , \label{eq:fsum} 
\end{align}
where $e$ is the charge of an electron, $\varepsilon_0$ the free-space permittivity, and $m_e$ the electron rest mass. This sum rule arises as an application of the KK relation of \eqref{kk}: at high enough frequencies $\omega$, the material must be nearly transparent, with only a perturbative term that arises from the individual electrons without any multiple-scattering effects. The sum rule of \eqref{fsum} is the critical constraint on refractive index: intuitively, it places a limit on the distribution of oscillators in any material; mathematically, it limits the distribution of the measure $\omega' \Im \chi(\omega') {\rm d}\omega'$ that appears in \eqref{kk}. 

To simulate any possible material, we must discretize \eqreftwo{kk}{fsum} in a finite-dimensional basis. If we use a finite number $N$ of localized basis functions (e.g. a collocation scheme~\cite{boyd2001chebyshev} of delta functions), straightforward insertion of the basis functions into \eqref{fsum}, in tandem with the constraint of \eqref{kk}, leads to a simple representation of the susceptibility:
\begin{align}
    \Re \chi(\omega) &= \sum_{i=1}^N  \frac{c_i \omega_p^2}{\omega_i^2 - \omega^2},  \label{eq:chiosc} \\
    \sum_i c_i &= 1.
    \label{eq:sumosc} 
\end{align}			 
\Eqref{chiosc} distills the Kramers--Kronig relation to a set of ``lossless" Drude--Lorentz oscillators with transition frequencies $\omega_i$ and relative weights, or oscillator strengths, $c_i$. \Eqref{sumosc} is a renormalized version of the $f$-sum rule of \eqref{fsum}, thanks to the inclusion of $\omega_p^2$ in the numerator of \eqref{chiosc}. There is one more important restriction on the $c_i$ values: they must all be positive, since $\omega' \Im \chi(\omega')$ must be positive for a passive material (under an $e^{-i \omega t}$ time-harmonic convention). Given \eqreftwo{chiosc}{sumosc}, it now becomes plausible that there is a bound on refractive index: the oscillators of \eqref{chiosc} represent all possible lineshapes, and the sum rule of \eqref{sumosc} restrict the oscillator strengths, and effective plasma frequencies, of the constituent oscillators. 

It is important to emphasize that the constants in the sum rule of \eqref{fsum} are indeed constants; in particular, that the mass $m_e$ is the free-electron mass and not an effective mass of an electron quasiparticle. In interband models~\cite{Cardona2005,Kaxiras2019}, the linear susceptibility can be written as a sum of Drude--Lorentz oscillators similar to \eqref{chiosc} and containing the effective masses of the relevant bands. But for those models, the sum over all bands leads to the free-electron mass in the final sum rule~\cite{Kaxiras2019}. Alternatively, one can use the fact that electrons can be considered as free, non-interacting particles in the high-frequency limit~\cite{Landau2013}. Thus the only variable in the sum rule is the electron density, which itself does not vary all that much over all relevant materials at standard temperatures and pressures. It is equally important to emphasize that the representation of \eqref{chiosc} does not rely on any of the standard assumptions of interband models (no many-body effects, periodic lattice, etc.), and is valid for \emph{any} linear (isotropic) susceptibility, assuming only causality. \Eqref{chiosc} is \emph{not} a Drude--Lorentz approximation or model; instead, it is a first-principles representation of the Kramers--Kronig relations.

To determine the maximum possible refractive index, one could maximize \eqref{chiosc} over all possible sets of parameter values for the oscillator strengths and transition frequencies, $c_i$ and $\omega_i$, respectively. However, a global optimization over the Drude--Lorentz form that is nonlinear in the $\omega_i$ will be practically infeasible for a large set of transition frequencies. Instead, we \emph{a priori} fix a very large number of possible oscillator transition frequencies $\omega_i$, and then treat only the corresponding oscillator strengths $c_i$ as the independent degrees of freedom. This ``lifting'' transforms a moderately large nonlinear problem to a very large linear one, and there are well-developed tools for rapidly solving for the global optima of linear problems~\cite{Nocedal2006,Vanderbei2015}. 

Crucially, not only is the susceptibility linear in the oscillator-strength degrees of freedom $c_i$, but so are many possible quantities of interest for constraints: first-, second-, and any-order frequency derivatives of the susceptibility, loss rates (the imaginary part of the susceptibility), etc. Thus maximizing refractive index over any bandwidth, or collection of frequency points, subject to any constraints over bandwidth or dispersion, naturally leads to generic linear programs of the form:
\begin{equation}
    \begin{aligned}
        & \underset{c}{\text{maximize}} & &  z^T c \\
        & \text{subject to}       & & A_j^T c + b_j \leq 0 \\
         & & & 1^T c = 1 \\
         & & & c \geq 0. \label {eq:opt_gen}
    \end{aligned}
\end{equation} 
where $c$ without a subscript denotes the length-$N$ vector comprising the oscillator strengths, $j$ indexes any number of possible constraints, the constraint $1^T c = 1$ corresponds to the sum rule $\sum_i c_i = 1$, and $z$, $A_i$, and $b_i$ are the appropriate vectors and matrices that are determined by the specific objectives, constraints, and frequencies of interest. There are well-developed tools for rapidly solving for the global optima of linear problems such as \eqref{opt_gen}, and in the following sections we identify important questions that take this form.

\Eqref{opt_gen} represents the culmination of our transformation of generic refractive-index-maximization problems to linear programs. A natural question might be why we use the Kramers--Kronig relation, \eqref{kk}, and sum rule, \eqref{fsum}, for material susceptibility $\chi$ instead of refractive index $n$ directly? In fact, one could replace all of the preceding equations with their analogous refractive-index counterparts, and arrive at an analogous linear-program formulation for refractive index. But the bounds would be significantly looser, the physical origins for which we explain in \secref{nKK}. Instead, it turns out that the susceptibility-based formulation presented above leads to bounds that are rather tight. 

\begin{figure} [t!]
    \includegraphics[width=1\linewidth]{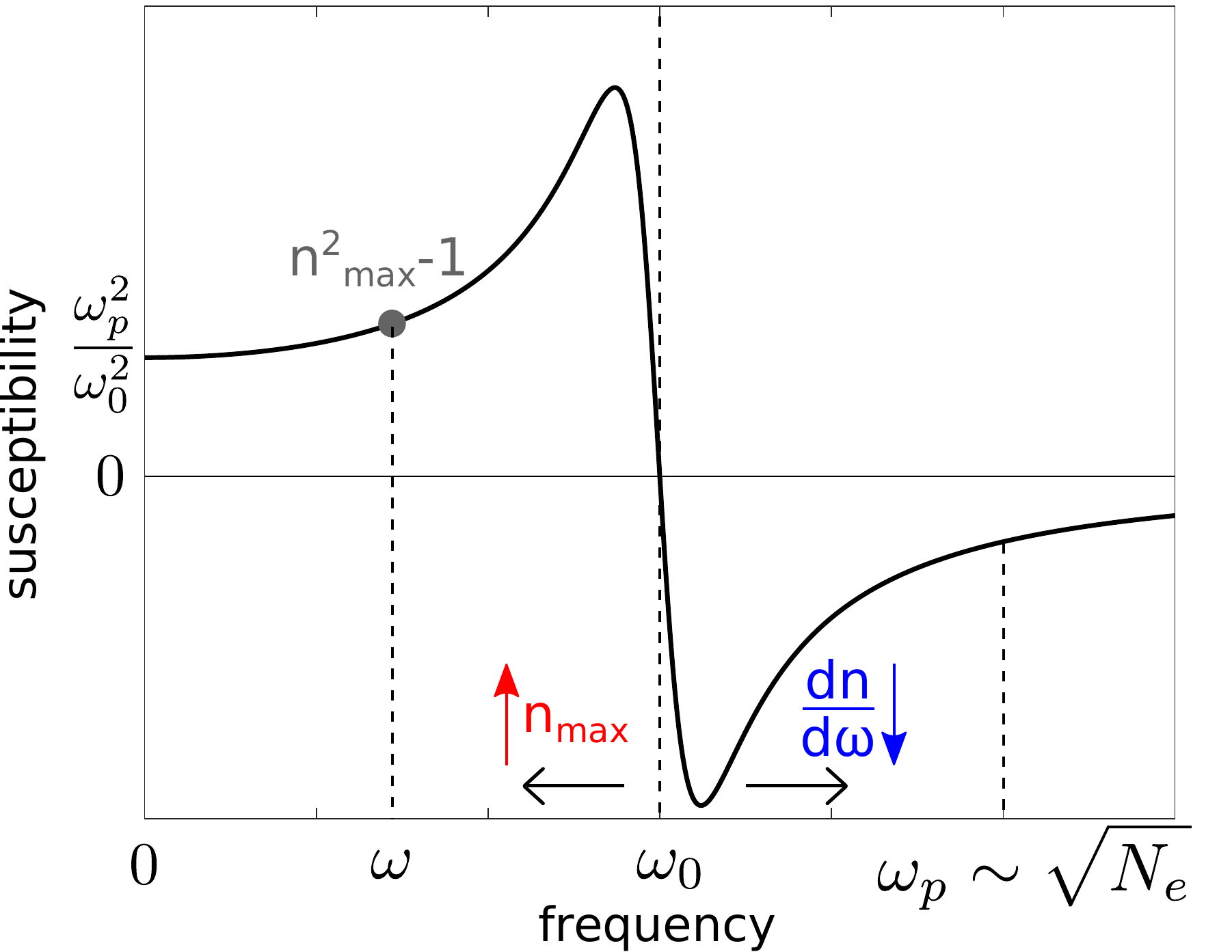} 
    \caption{Schematic representation of a single Drude--Lorentz oscillator, depicting the tradeoff between refractive index and dispersion. Decreasing the resonance frequency $\omega_0$ increases the ratio $\frac{\omega_p^2}{\omega_0^2}$ and hence the maximum refractive index $n_{\textrm{max}}$ at $\omega$, but at the cost of higher dispersion $\frac{{\rm d} n}{{\rm d} \omega}$ (and vice versa for increasing $\omega_0$). The plasma frequency $\omega_p = \sqrt{\frac{N_e e^2}{\varepsilon_0 m_e}}$ is determined by the material's electron density.}  
    \label{fig:singleosc} 
\end{figure} 		

\section{Single-frequency bound} \label{sec:sf}
\subsection{Fundamental limit}
A canonical version of the refractive-index question is: what is the largest possible refractive index of a transparent (lossless) medium, at frequency $\omega$, subject to some maximum allowable dispersion? The dispersion constraint is important for many applications, from metalenses to photovoltaics, where one may want to operate over a reasonable bandwidth or minimize the phase- and/or group-velocity variability that can be difficult to overcome purely by design~\cite{Genevet2017,Chen2018,Chung2020}. Given the susceptibility representation of \eqreftwo{chiosc}{sumosc} in \secref{linprog}, we can formulate the maximum-refractive-index question in terms of the susceptibility, and then transform the optimal solution to a bound on refractive index. We assume here a nonmagnetic medium, in which case we can connect electric susceptibility to refractive index; in \secref{mag} we show that the same bounds apply even in the presence of magnetic response.

The formulation of this canonical single-frequency refractive-index maximization as a linear program is straightforward. The Kramers--Kronig representation of \eqref{chiosc} can be written as $\chi(\omega) = \sum_i c_i f_i(\omega)$, where $f_i$ at frequency $\omega$ is given by $f_i(\omega) = \omega_p^2 / (\omega_i^2 - \omega^2)$ and $\chi(\omega)$ is linear in the $c_i$ values. The dispersion, as measured by the frequency derivative of the real part of the susceptibility, has the same representation but with $f_i(\omega)$ replaced by its derivative $f_i'(\omega) = 2\omega \omega_p^2 / (\omega_i^2 - \omega^2)^2$. Then, the largest possible susceptibility at frequency $\omega$, with dispersion constrained to be smaller than an application-specific constant $\chi'$, is the solution of the optimization problem:
\begin{equation}
    \begin{aligned}
        & \underset{c_i}{\text{maximize}} & &  \Re \chi(\omega) = \sum_{i=1}^N c_i \, f_i(\omega) \\
        & \text{subject to}       & & \frac{{\rm d} \Re \chi(\omega)}{{\rm d} \omega} =  \sum_{i=1}^N c_i \, f_i'(\omega) \leq \chi', \\
        & & &  \sum_{i=1}^N c_i  = 1, \ c_i \geq 0. \label {eq:opt_sf}
    \end{aligned}
\end{equation} 
\Eqref{opt_sf} is of the general linear-program form in \eqref{opt_gen}. Comparing the two expressions, the vector $z$ has $f_i(\omega)$ as its elements. There is only a single index $j$, with matrix $A_1$ given by a single column with values $f_i'(\omega)$ and $b_1$ is a vector of 1's multiplied by $-\chi'$. To computationally optimize the maximum-index problem of \eqref{opt_sf}, one must simply represent a sufficiently large space of possible oscillator frequencies $\omega_i$. Since we are interested in transparent (lossless) media, there should not be any oscillator at the frequency of interest $\omega$ (otherwise there will be significant absorption). Nor should there be any oscillator frequencies $\omega_i < \omega$, which can only reduce the susceptibility at $\omega$. Thus, one only needs to consider oscillator strengths $\omega_i$ greater than $\omega$. 

Strikingly, for any frequency $\omega$, electron density $N_e$ (or plasma frequency $\omega_p$), and allowable dispersion $\chi'$, the optimal solution to \eqref{opt_sf} is always represented by a single nonzero oscillator, with strength $c_0 = 1$ and frequency $\omega_0 = \omega \sqrt{1 + \sqrt{2 \omega_p^2 / (\omega^3 \chi')}}$. We prove in the {\SM} that this single-oscillator solution is globally optimal. The intuition behind the optimality of a single oscillator can be understood from \figref{singleosc}. The susceptibility of a single oscillator is governed by three frequencies: the frequency of interest, $\omega$, the oscillator frequency, $\omega_0$, and the electron-density-based plasma frequency, $\omega_p$. The static susceptibility of such an oscillator at zero frequency is given by $\chi = \omega_p^2 / \omega^2$. This sets a starting point for the susceptibility that ideally should be as large as possible. The plasma frequency is fixed for a given electron density, and thus the only way to increase the static susceptibility is to reduce the oscillator frequency $\omega_0$ (as indicated by the black left arrow). Yet this comes with a tradeoff: as $\omega_0$ decreases, the oscillator nears the frequency of interest, and the dispersion naturally increases. Hence for minimal dispersion one would want as large of an oscillator frequency as possible. A constraint on allowable dispersion thus imposes a bound on how small of an oscillator frequency one can have, and the maximum refractive index is achieved by concentrating all of the available oscillator strength, determined by the f-sum rule, at that frequency.

The single-oscillator optimality of the solution to \eqref{opt_sf} leads to an \emph{analytical} bound on the maximum achievable refractive index. Denoting a maximal refractive-index dispersion $n' = \chi' / 2n$ (from $\chi = n^2 - 1$), straightforward algebra (cf. {\SM}) leads to a general bound on achievable refractive index:
\begin{align}
\frac{(n^2-1)^2}{n} \leq \frac{\omega_p^2 n'}{\omega}.  \label{eq:nbound}
\end{align}
\Eqref{nbound} is a key result of our paper, delineating the largest achievable refractive index at any frequency for any passive, linear, isotropic material. \Eqref{nbound} highlights the three key determinants of maximum refractive index: electron density, allowable dispersion, and frequency of interest. We will discuss each of these three dependencies in depth. First, though, there is a notable simplification of the refractive-index bound, \eqref{nbound}, when the refractive index is moderately large. In that case, the left-hand side of \eqref{nbound} is simply the cube of $n$; taking the cube root, we have the high-index ($n^2 \gg 1$) bound:
\begin{align}
n \leq \left( \frac{\omega_p^2 n'}{\omega} \right)^{1/3}.  \label{eq:nbound_high}
\end{align}
The cube-root dependence of the high-index bound, \eqref{nbound_high} is a strong constraint: it says that increasing electron density or allowable dispersion by even a factor of 2 will only result in a $\sqrt[3]{2} \approx 1.26\textrm{X}$ enhancement. Similarly, even an order-of-magnitude, 10X increase in either variable can only enhance refractive index by a little more than 2X. Thus the opportunity for significant increases in refractive index are highly limited. The cube-root dependence that is responsible for this constraint is new and surprising: conventional arguments suggest that refractive index should scale with the square root of electron density~\cite{Landau2013}. Moreover, applying our analysis to the Kramers--Kronig representation of refractive index also leads to square-root scaling. It is the fact that the \emph{susceptibilities} of nonmagnetic materials, in addition to their refractive indices, must satisfy Kramers--Kronig relations, that ultimately leads to the tighter cube-root dependence, as further discussed in \secref{nKK}.
\begin{figure*} [t!]
    \includegraphics[width=1\linewidth]{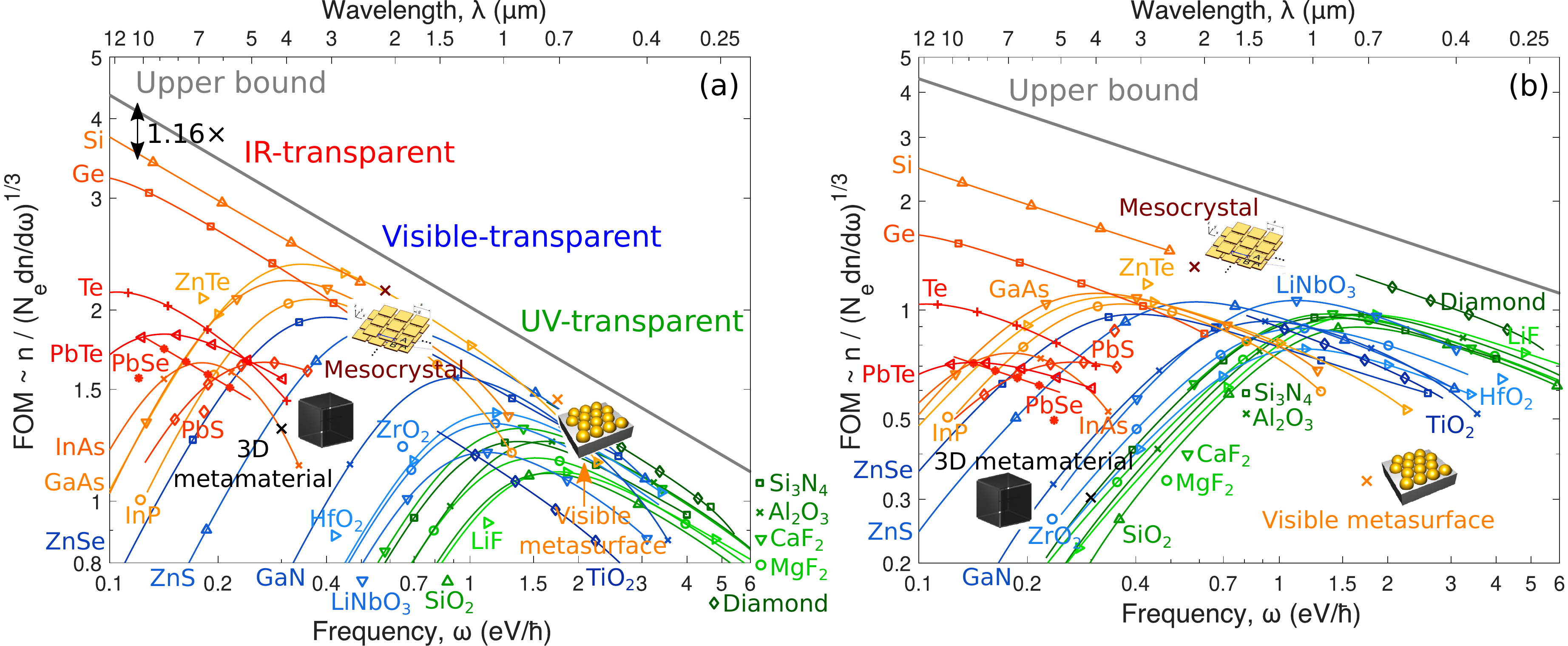} 
    \caption{Comparison of representative high-index materials, as well as three metamaterial designs (visible metasurface~\cite{Kim2016}, 3D metamaterial~\cite{Shin2009}, mesocrystal~\cite{Chang2016}) from the literature, as measured by the FOM = $\left[ \frac{(n^2-1)^2}{n\omega_p^2 n'} \right]^{1/3} \approx \frac{n}{N_e n'}$ for $n\gg1$ ($N_e n'$ normalized to that of valence SiO$_2$ at 400 nm), plotted against the material-independent bound in \eqref{metric}. Shown above are two figures, based on (a) total and (b) valence electron density, which only shifts the FOM for each material without distorting qualitative features. The materials can be broadly classified into three categories depending on the spectrum at which they are transparent---UV (LiF~\cite{Moore1982}, MgF\ts{2}~\cite{Moore1982}, CaF\ts{2}~\cite{Moore1982}, SiO\ts{2}~\cite{Malitson1965a,Tan1998},  Al\ts{2}O\ts{3}~\cite{Malitson1965a,Weber1986}, Si\ts{3}N\ts{4}~\cite{Luke2015}, diamond~\cite{Peter1923}), visible (HfO\ts{2}~\cite{Wood1990}, ZrO\ts{2}~\cite{Wood1982a}, LiNbO\ts{3}~\cite{Zelmon1997a}, ZnS~\cite{Debenham1984a,Klein1986}, GaN~\cite{Barker1973}, ZnSe~\cite{Amotchkina2020},  TiO\ts{2}~\cite{Devore1951}), and IR (ZnTe~\cite{Li1984a}, InP~\cite{Pettit1965a}, GaAs~\cite{Skauli2003a}, Si~\cite{Chandler2005}, InAs~\cite{Lorimor1965a}, Ge~\cite{Burnett2016a}, PbS~\cite{Zemel1965}, PbSe~\cite{Zemel1965}, Te~\cite{Caldwell1959,Bhar1976a}, PbTe~\cite{Weiting1990}). Most materials approach the bound quite closely at different frequencies, with silicon at IR outperformed only by a factor of 1.16 relative to the bound. The parabolic shape bending downwards at IR frequencies can be attributed to phonon dispersion, notable exceptions being silicon and germanium, which are IR-inactive and hence highly transparent even in mid-IR (i.e. very non-dispersive).}  
    \label{fig:dispbound} 
\end{figure*} 

To investigate the validity of our bounds of \eqreftwo{nbound}{nbound_high}, we compare them to the actual refractive indices of a wide range of real materials. To compare the bound to a real material at varying frequencies, we must account for the different electron densities, dispersion values, and frequencies of interest for those materials. To unify the comparisons, we use the bound of \eqref{nbound} to define a material-dependent refractive-index ``figure of merit,''
\begin{align}
\textrm{FOM} \equiv \left[ \frac{(n^2-1)^2}{n\omega_p^2 n'} \right]^{1/3} \leq \frac{1}{\omega^{1/3}}.  \label{eq:metric}
\end{align}
which is approximately the refractive index rescaled by powers of the plasma frequency and allowable dispersion. On the right-hand side of \eqref{metric} is the factor $1/\omega^{1/3}$, which is the upper bound to the material figure of merit for any material.

\Figref{dispbound} compares the material-figure-of-merit bound (solid black line) to the actual material figure of merit for a wide range of materials (colored lines and markers)~\cite{Moore1982,Malitson1965a,Tan1998,Weber1986,Luke2015,Peter1923,Wood1990,Wood1982a,Zelmon1997a,Debenham1984a,Klein1986,Barker1973,Amotchkina2020,Devore1951,Li1984a,Pettit1965a,Skauli2003a,Chandler2005,Lorimor1965a,Burnett2016a,Zemel1965,Caldwell1959,Bhar1976a,Weiting1990}. We use the experimentally determined refractive indices and dispersion values for each material. Parts (a) and (b) of the figure are identical except for the values of the electron density: in part (b), we use the total electron density of each material, while in part (a) we use only the valence electron density. The valence-electron-density bound is not a rigorous bound, but in practice it is only the valence electrons that contribute to the refractive index at optical frequencies, and one can see that the bound in (a) is tighter than that of (b) due to the use of the valence densities, while not being surpassed by any real materials. Within the materials considered, we use the line and marker colors to distinguish materials that are transparent at infrared (IR), visible, and ultraviolet (UV) frequencies, respectively. The higher the frequency of interest, the lower the material FOM bound is (and the lower the refractive-index bound is), because at higher frequencies the oscillator frequency must increase to prevent the dispersion value from surpassing its limit, and a higher oscillator frequency reduces the electrostatic index that sets a baseline for its ultimate value (as can be seen in \figref{singleosc}).

Three metamaterials structures~\cite{Shin2009, Kim2016, Chang2016} are included in \figref{dispbound}. These metamaterials are patterned to exhibit anomalously large effective indices (ranging from 5 to 10). Ultimately, these metamaterials are configurations of electrons that effectively respond as a homogeneous medium with some refractive index, and thus they too are subject to the bounds of \eqreftwo{nbound}{nbound_high}. Indeed, as shown in \figref{dispbound}, two of the metamaterials approach the valence-electron bound line, but do not surpass it. Their high refractive indices are accompanied by dramatically increased chromatic dispersion.

Many materials can approach the bound over a small window of frequencies where their dispersion is minimal relative to their refractive index. Two outliers are silicon and germanium, which approach the bound across almost all frequencies at which they are transparent. Silicon, for example, has a refractive index ($n=3.4$) that is within 16\% of its valence-density-based limit. The key factor underlying their standout performance is a subtle one: the absence of optically active phonon modes. It turns out that optical phonons primarily increase the dispersion of a material's refractive index without increasing its magnitude. From a bound perspective, this can be understood from the sum rule of \eqref{fsum}. In that sum rule, the total oscillator strength is connected to the electron density of a material, divided by the free-electron mass. Technically, there are additionally terms in the sum rule for the protons and neutrons~\cite{Yan1995}. However, their respective masses are so much larger than those of electrons that their relative contributions to the sum rule are insignificant. Similarly, because phonons are excitations of the lattice, their contribution to refractive index comes from the proton and neutron sum-rule contributions, and are necessarily insignificant in magnitude at optical frequencies. They can, however, substantially alter the dispersion of the material, and indeed that is quite apparent in the refractive indices of many of the other materials (e.g. GaAs, InP, etc.), which thus tend to fall short of the bounds at many frequencies. This result suggests that ideal high-index materials should not host active optical phonons, which increase dispersion without increasing refractive index.

\begin{table*}
\begin{center}
\begin{tabular} { |c|c|c|c|c| }
\hline
 Material & Electron density  &Dispersion $\frac{{\rm d} n}{{\rm d} \omega}$ (eV$^{-1}$) &Refractive index n &Bound on n\\ \cline{3-5}

~ & $N_e$ (10$^{23}$ cm$^{-3}$) &  \multicolumn{3}{c|}{ (averaged over 400--700 nm) } \\
 
\hline
MgF$_2$   & 4.85 & 0.0059 & 1.38 & 1.58 \\
\hline
CaF$_2$   & 3.92 & 0.0076 & 1.43 & 1.60  \\
\hline
SiO$_2$   & 4.25 & 0.0112 & 1.46 & 1.73 \\
\hline
Al$_2$O$_3$  & 5.67 & 0.0176 & 1.77 & 2.04 \\
\hline
Si$_3$N$_4$  & 4.39 & 0.0514 & 2.06 & 2.48 \\
\hline
HfO$_2$  & 4.65 & 0.0482 & 2.13 & 2.49 \\
\hline
ZrO$_2$   & 4.75 & 0.0597 & 2.18 & 2.63 \\
\hline
LiNbO$_3$ & 4.52 & 0.1266 & 2.34 & 3.12 \\
\hline
C (diamond) & 7.04 & 0.0436 & 2.43 & 2.74 \\
\hline
GaN  & 3.03 & 0.1448 & 2.45 & 2.97 \\
\hline
TiO$_2$  & 5.11 & 0.3342 & 2.72 & 4.17 \\
\hline
    Metamaterial \footnote{refers to the metamaterial in \citeasnoun{Kim2016}, here evaluated at $\approx$ \SI{710}{nm}.} & 0.59 & $\approx$ 4.1 & $\approx$ 5.1 & $\approx$ 5.7 \\
\hline
\end{tabular}
\end{center}
\caption{High-index materials transparent over the visible spectrum, showing the (valence) electron density $N_e$, dispersion $\frac{{\rm d} n}{{\rm d} \omega}$, refractive index $n$, and upper bound on $n$ for each material. The table shows that refractive index, as well as its bound, increases with dispersion, and that they closely approach the bound. Except for the metamaterial, all the quantities listed above are averaged over 400--700 nm.}
\label{visible_highn}
\end{table*}
		  		    
Table~\ref{visible_highn} presents numerical values of valence electron densities, dispersion values, refractive indices, and their bounds for representative materials averaged over the visible spectrum (see {\SM} for more details on bounds for nonzero bandwidth). One can see that for a wide variety of materials~\cite{Moore1982,Malitson1965a,Tan1998,Weber1986,Luke2015,Peter1923,Wood1990,Wood1982a,Zelmon1997a,Barker1973,Devore1951} and dispersion values, there is a close correspondence between the actual refractive index and the bound, for both natural materials and artificial metamaterials. Taken together, \figref{dispbound} and Table~\ref{visible_highn} show that many materials can closely approach their respective bounds, showing little room for improvement at the dispersion values naturally available. These results also cast doubt about the result of \citeasnoun{DiMei2018}: a refractive index of 26 at optical frequencies is an order of magnitude larger than any of the natural materials in Table~\ref{visible_highn}. Because of the cube-root scaling of the bound of \eqreftwo{nbound}{nbound_high}, a 10X increase in refractive index requires a 1000-fold increase in electron density or dispersion. Large dispersion would inhibit the possibility for the broadband nature of the result in \citeasnoun{DiMei2018}, hence the only remaining possibility is a $\approx 1000X$ increase in electron density. Yet this would be orders of magnitude larger than the largest known electron densities~\cite{Kittel1996}. Hence, our results strongly suggest that the light-bending phenomena of \citeasnoun{DiMei2018} are due to diffractive or nonlinear effects, instead of a linear refractive index.

\subsection{Maximum index versus chromatic dispersion}
\begin{figure} [t!]
    \includegraphics[width=1\linewidth]{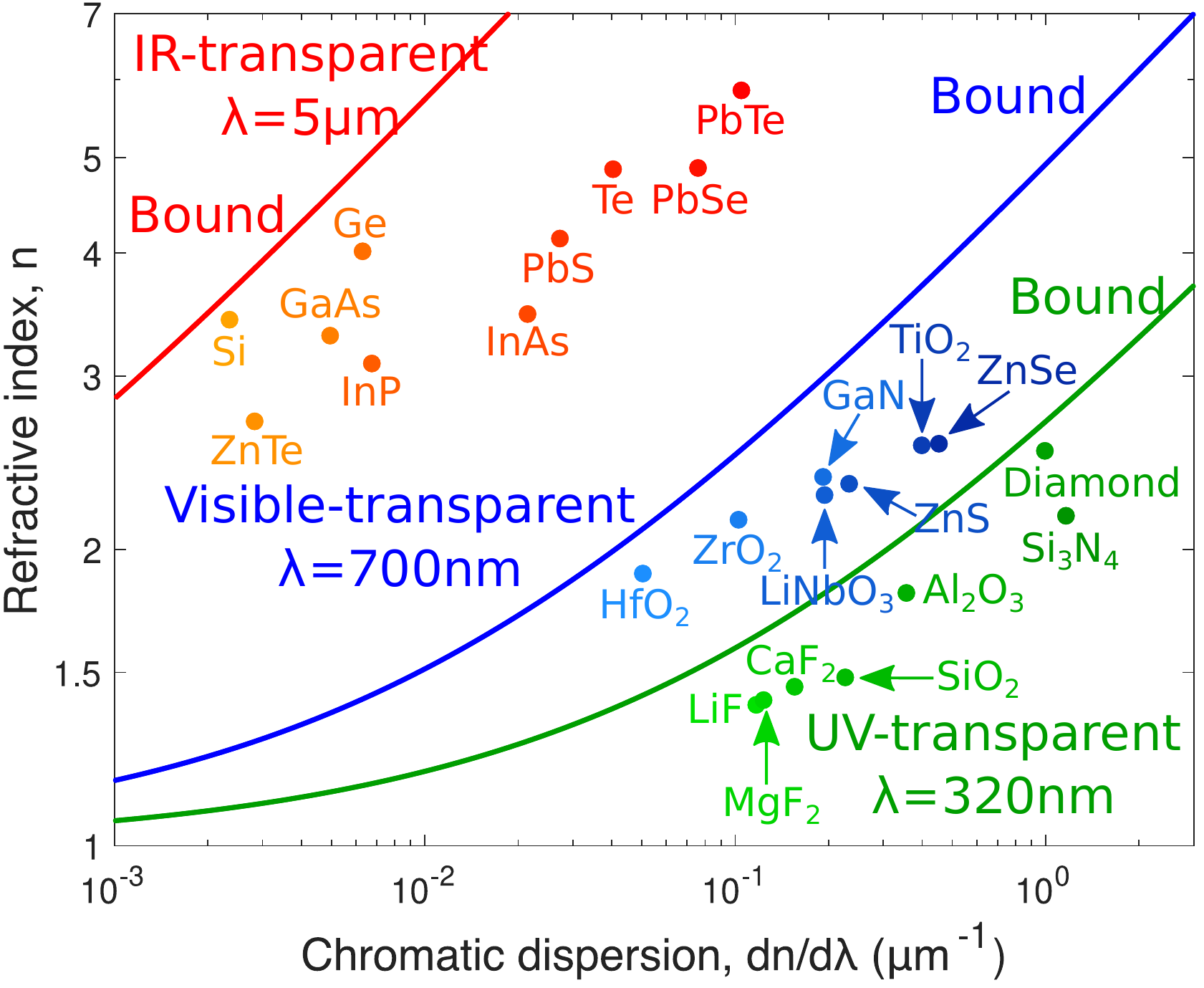} 
    \caption{Refractive indices of various materials evaluated at three different wavelengths (320 nm, 500 nm, 5 $\mu$m) based on the spectrum at which they are transparent (UV, visible, and IR respectively), compared to their respective bounds. For each wavelength, the average electron density of all materials belonging to that wavelength was used to compute the bound. The bounds show refractive index increasing with dispersion, as measured by (the magnitude of) chromatic dispersion $\frac{{\rm d} n}{{\rm d} \lambda}$, as also demonstrated by materials closely approaching the bounds.}
    \label{fig:rivsdisp} 
\end{figure}
\begin{figure*}[bth]
    \includegraphics[width=1\linewidth]{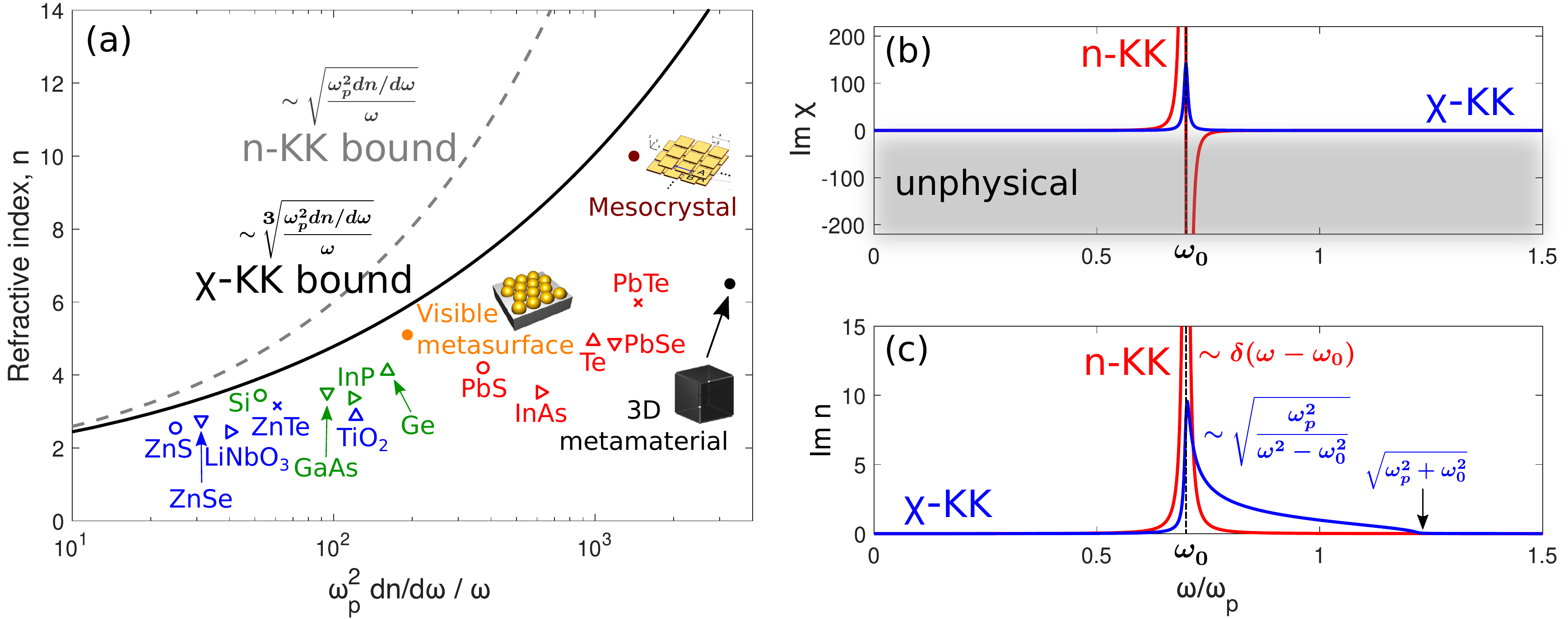} 
    \caption{(a) Comparison of bounds based on Kramers--Kronig relations on susceptibility, \eqref{nbound}, and refractive index, \eqref{nkkbound}, denoted as $n$-KK and $\chi$-KK bound respectively. Natural materials are categorized in terms of the frequencies at which they obtain the highest refractive index (visible, near-IR, and mid-IR, marked as blue, green, and red respectively). All the materials lie below the $\chi$-KK bound. (b) Optimal $\Im \chi$ profiles attaining the $n$-KK and $\chi$-KK bound. Around the resonance frequency $\omega_0$, $\Im \chi$ for the former goes negative, which is not allowed by passivity. (c) Optimal $\Im n$ profiles attaining the $n$-KK and $\chi$-KK bound. In contrast to an infinitely sharp resonance for the former, the latter is characterized by a broadened lineshape to the right of $\omega_0$. For (b) and (c), the loss rate was taken to be small but nonzero ($\gamma = 0.01\omega_p$) for purposes of illustration.}  
    \label{fig:nkk} 
\end{figure*} 
To visualize the tradeoff between maximal refractive index and dispersion, \figref{rivsdisp} depicts the refractive-index bound of \eqref{nbound} as a function of chromatic dispersion, for materials transparent at three different wavelengths: infrared ($\lambda = \SI{5}{um}$), visible ($\lambda = \SI{700}{nm}$), and ultraviolet ($\lambda = \SI{320}{nm}$). We use dispersion with respect to wavelength, i.e. $dn/d\lambda$, instead of frequency, as the wavelength derivative is commonly used in optics~\cite{Born2013}. Without careful attention to the wavelength of interest, it would appear that refractive index tends to decrease as dispersion increases: silicon, for example, has both a higher refractive index and smaller dispersion than titanium dioxide, at their respective transparency wavelengths. Yet our bound of \eqref{nbound} highlights the key role that wavelength is playing in this comparison: the bound shows that maximum index must decrease with increased dispersion but increase at longer wavelengths. Within each color family in \figref{rivsdisp}, wavelength is held constant, and then it is readily apparent that maximum index increases as a function of chromatic dispersion. One can see that in each wavelength range, many materials are able to approach our bounds across a wide range of dispersion levels. The largest gaps between actual index and that of the bound occur for infrared III-V and II-VI materials, due to the presence of active optical phonons, as discussed above.

At visible and UV frequencies, where phonon contributions are negligible, the deviation of refractive indices from their respective bounds can be attributed to the distributions of oscillator strengths, manifest in the frequency dependence of $\Im \chi(\omega)$. The larger the frequency spread (variance) of $\Im \chi$ relative to its average frequency, the more a material's refractive index falls short of the bound (cf. {\SM}). Note that for a fixed frequency of interest, the average frequency of the optimal oscillator depends directly on the maximum allowable dispersion: larger dispersion implies smaller oscillator frequency, and vice versa. Hence, higher-dispersion materials have smaller average oscillator frequencies, which reduces the total variance allowed before significant reductions relative to the bounds arise. Diamond would appear to be an exception, but that is only because its valence electron density is much larger than average; its gap to its respective bound is as expected. To summarize: highly dispersive materials are more sensitive to deviations of $\Im \chi(\omega)$ from the ideal delta function than are small-dispersion materials. A direct comparison can be done for TiO$_2$ and HfO$_2$, which have similar oscillator spreads but a smaller center frequency for TiO$_2$. This explains why TiO$_2$ is farther from its bound than is HfO$_2$, and explains the general trend of increasing gaps with increasing dispersion.


\subsection{Bounds from refractive-index KK relations}
\label{sec:nKK}
In \secref{linprog}, we noted the importance of using Kramers--Kronig relations for the \emph{susceptibility} instead of KK relations for refractive index. Here, we briefly show the bound that can be derived via refractive-index KK relations, and explain why the two bounds are quite different.

Analogous to the sum rule of \eqref{fsum}, there is a sum rule on the distribution of the imaginary part of refractive index that also scales with the electron density: $\int_0^{\infty} \omega \Im n(\omega) \,{\rm d}\omega = \pi \omega_p^2/4$ (\citeasnoun{King1976}). Similarly, there is a KK relation for refractive index that exactly mimics \eqref{kk}. Together, following the same mathematical formulation as in \secref{linprog}, one can derive a corresponding bound on refractive index given by (cf. {\SM}):
\begin{align}
n \leq 1 + \frac{\omega_p}{2} \sqrt{\frac{n'}{\omega}}.  \label{eq:nkkbound}
\end{align}		
To distinguish the two bounds from each other, we will denote this bound, \eqref{nkkbound}, as the $n$-KK bound, and the susceptibility-based bound, \eqref{nbound}, as the $\chi$-KK bound.  \Eqref{nkkbound} shows that the $n$-KK bound has a square-root dependence on the parameter $\frac{\omega_p^2 n'}{ \omega}$, in contrast to the cube-root dependence for the $\chi$-KK bound (explicitly shown in \eqref{nbound_high} for high-index materials). The $n$-KK bound is always larger than the $\chi$-KK bound (cf. {\SM}), and the square-root versus cube-root dependencies implies that the gap increases with dispersion and electron density. \Figref{nkk}(a) shows the difference between the two bounds, and the increasing gap between them at large plasma frequencies or allowable dispersion. \Figref{nkk}(b,c) shows the physical origins of the discrepancy between the two approaches. The optimal $n$-KK solution has a delta-function imaginary part of its refractive index, as in \figref{nkk}(c), concentrating all of the imaginary part in a single refractive-index oscillator. Yet for a delta function in $\Im n$, the imaginary part of the electric susceptibility \emph{must} go negative in a nonmagnetic material, as in \figref{nkk}(b), which is unphysical in a passive medium. (At this point, one might wonder if the n-KK bound is achievable by allowing for magnetic response. However, as shown in \secref{mag}, a non-zero magnetic susceptibility will not help in overcoming the $\chi$-KK bound, as the overall material response under the action of an electromagnetic field is still bound by the $f$-sum rule in \eqref{fsum}.) By contrast, the optimal solution in the $\chi$-KK bound is a delta function in susceptibility, as in \figref{nkk}(b), which yields a smoother, physical distribution of $\Im n(\omega)$, as seen in \figref{nkk}(c). Hence another way of understanding the surprising cube-root dependency of our bound is that it arises as a unique consequence of the fact that both refractive index $n$ and its square, $n^2 = \chi + 1$, obey Kramers--Kronig relations~\cite{Nussenzveig1972}.
\begin{figure*}[tbh]
    \includegraphics[width=1\linewidth]{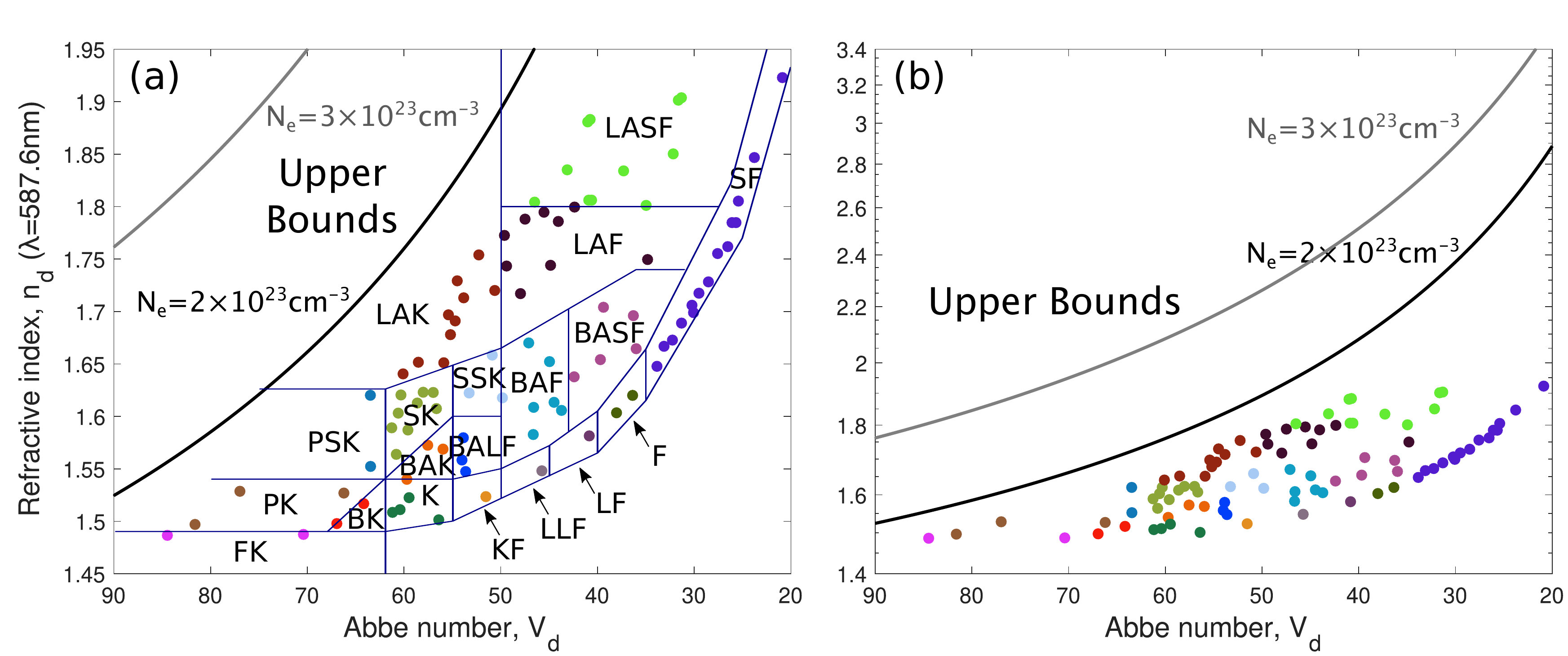} 
    \caption{(a) Abbe diagram showing the glasses categorized depending on their refractive indices (at 587.6 nm) $n_d$ and Abbe number $V_d$, compared to the bounds for electron density $N_e = 2 \times 10^{23}$ cm$^{-3}$ (black line) and $N_e = 3 \times 10^{23}$ cm$^{-3}$ (gray line). (b) Same plot but shown in logarithmic scale with larger range of values to fully illustrate the bounds. The data for different glass categories was obtained from Ref.~\cite{Schott2011}.} 
    \label{fig:abbe} 
\end{figure*}

\section{Bound on optical glasses} \label{sec:abbe}
The optical glass industry has put significant effort into designing high-index, low-dispersion optical glasses. Thus, transparent optical glasses provide a natural opportunity to test our bounds. It is common practice to categorize refractive indices at specific, standardized wavelengths. The refractive index $n_d$ refers to refractive index at the Fraunhofer d spectral line~\cite{Hearnshaw1990}, for wavelength $\lambda = \SI{587.6}{nm}$, in the middle of the visible spectrum. Dispersion is measured by the Abbe number $V_d$~\cite{Hovestadt1902,Bergmann1999}:
\begin{align}
V_d = \frac{n_d-1}{n_F - n_C},  \label{eq:abbedef}
\end{align}
where $n_F$ and $n_C$ are evaluated at \SI{486.1}{nm} and \SI{656.3}{nm}, the Fraunhofer F and C spectral lines, respectively (the Abbe number can be defined differently based on other spectral lines, but the above convention is commonly used to compare optical glasses~\cite{Hartmann2010}). The quantity in \eqref{abbedef} cannot be directly constrained in our bound framework, as it is nonlinear in the susceptibility, but optical glasses of interest have sufficiently weak dispersion that their refractive indices can be approximated as linear across the visible spectrum. Then, we can relate the Abbe number directly to the dispersion of the material at $n_d$, ${\rm d}n/{\rm d}\omega$, and to the frequency bandwidth between the F and C spectral lines, $\Delta \omega_{FC}$:
\begin{align}
\frac{{\rm d} n}{{\rm d} \omega} \approx \frac{n_d-1}{\Delta \omega_{FC}} \frac{1}{V_d}. \label{eq:abbe}
\end{align}
which is valid for the wide range of glasses depicted in \figref{abbe} with up to only $5\%$ error. Inserting \eqref{abbe} into the refractive-index bound of \eqref{nbound}, we can write a bound on refractive index in terms of Abbe number $V_d$:
\begin{align}
\frac{(n_d^2-1)^2}{n_d(n_d-1)} \leq \frac{\omega_p^2}{\omega_d \Delta \omega_{FC}} \frac{1}{V_d}.  \label{eq:abbebound}
\end{align}

\Figref{abbe} plots the Abbe diagram~\cite{Hartmann2010} of many optical glasses along with our bounds for two representative electron densities, the valence electron density of silicon ($2 \times 10^{23}$ cm$^{-3}$) and the mean valence electron density of high-index materials shown in \figref{dispbound} ($3 \times 10^{23}$ cm$^{-3}$). From \figref{abbe}(a), there is a striking similarity of the shape of the upper bound and the trendlines for real optical glasses. Moreover, depending on the relevant electron density, the bounds may be quantitatively tight for the best optical glasses. \Figref{abbe}(b) zooms out and highlights the high-dispersion (large-Abbe-number) portion of the curve. The trend is very similar to that seen in \figref{dispbound} earlier: as dispersion increases, the gap between the bound and the refractive index of a real material increases, as the magnitude of the refractive index becomes more sensitive to broadening of the electron oscillator frequencies.

\section{Bandwidth-based bound} \label{sec:bw}
Instead of constraining the dispersion of a material refractive index, one might similarly require the refractive index to be high over some bandwidth of interest. A first formulation might be to maximize the average refractive index over some bandwidth, but this is ill-posed: an oscillator arbitrarily close to the frequency band of interest can drive the refractive index at the edge of the band arbitrarily high, and the average itself can also diverge. In any case, in a band of potentially large dispersion, the \emph{minimum} refractive index over the band is the more meaningful metric, as that will be the limiting factor in the desired optical response. Hence, maximizing the minimum refractive index over a bandwidth, i.e., solving a minimax problem, is the well-posed and physically relevant approach. We can pose the corresponding optimization problem problem for some bandwidth $\Delta\omega$ around a center frequency $\omega$ as:
\begin{equation}
    \begin{aligned}
        & \underset{c_i}{\text{maximize}} & &  \underset{\omega'}{\text{min}} \ \Re \chi(\omega') \qquad \omega-\frac{\Delta\omega}{2} \leq \omega' \leq \omega+\frac{\Delta\omega}{2} \\
        & \text{subject to}  & & \sum_{i=1}^N c_i  = 1, \ c_i \geq 0.  \label {eq:opt_bw}
    \end{aligned}
\end{equation}
where again we are considering only a transparency window in which the material is lossless. (As we show in the {\SM}, none of our bounds change substantially if small but nonzero losses are considered.) The solution to \eqref{opt_bw} is a single oscillator, analogous to the solution of \eqref{nbound}. In this case, the optimality conditions imply a single oscillator at the frequency $\omega + \Delta\omega/2$, i.e., exactly at the high-frequency edge of the band of interest. This optimal oscillator then implies a fundamental upper limit on the minimum refractive index over bandwidth $\Delta\omega$ around frequency $\omega$ to be ({\SM}):
\begin{align}
    n_{\textrm{min}} \leq \frac{\omega_p}{\sqrt{2\omega\Delta \omega}}.  \label{eq:nbound_loss}
\end{align}
\Eqref{nbound_loss} fundamentally constrains how large the minimal refractive index can be over any desired bandwidth. The only extra parameter is the material electron density, as encoded in $\omega_p$. The bound increases linearly as the square root of bandwidth $\Delta\omega$ decreases, which can be understood intuitively from the optimal refractive-index profile: decreasing the bandwidth $\Delta \omega$ effectively moves the infinitely sharp resonance (characterized by a delta-function $\Im \chi$) closer to the frequency of interest, thereby shifting the entire refractive-index spectrum upwards and resulting in higher $n_{\textrm{min}}$.

\begin{figure} [t!]
    \includegraphics[width=1\linewidth]{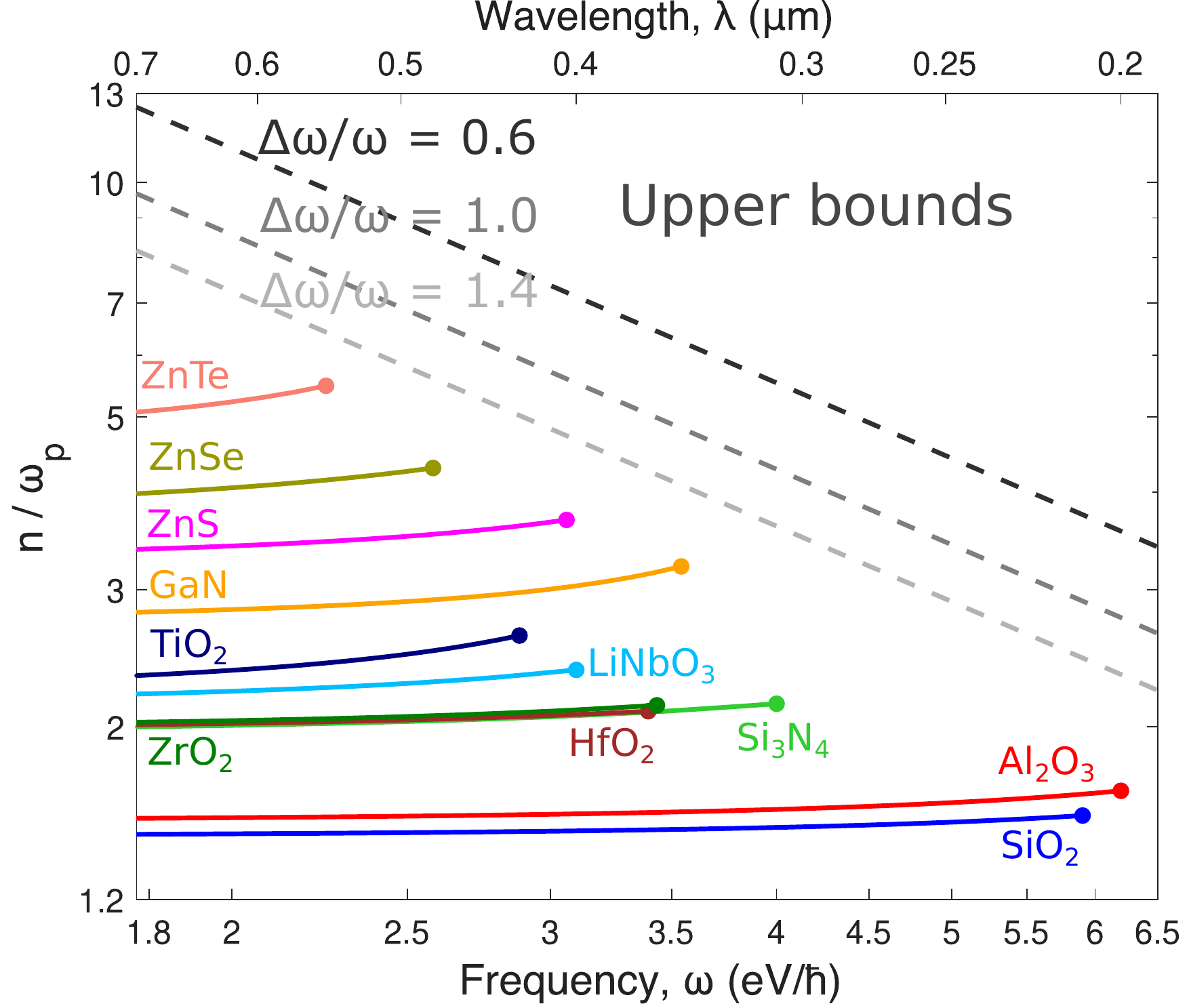} 
    \caption{Refractive index normalized by plasma frequency over the UV and visible spectrum, compared to the bounds for three different bandwidths. The dots for each material correspond to the edge of the transparency window, i.e. the frequency beyond which $\Im \chi$ becomes non-negligible. The plasma frequency for each material is normalized to that of SiO$_2$.}
    \label{fig:lossbound} 
\end{figure}

\Figref{lossbound} shows the refractive index, normalized by plasma frequency, for representative high-index materials in the visible and UV spectrum (over each of their transparency windows), compared to the bounds for three different bandwidths. Some materials like ZnTe and GaN more closely approach the bounds than others like TiO$_2$ and HfO$_2$, which can be traced back to their absorption loss ($\Im \chi$) spectrum. Ideally, the absorption is a delta function situated infinitesimally to the right of the transparency windows for each material, leading to a diverging refractive index at the edge of the transparency window (i.e. the dots for each curve). However, real materials are characterized by broad, smeared-out $\Im \chi$ and thus deviate from the ideal, single Drude--Lorentz response with infinitesimal loss rate. How much each material falls short of the bound signifies to what extent their $\Im \chi$ spectrum is, on average, concentrated away from the frequency of interest. One can deduce from \figref{lossbound} that, for example, ZnTe is characterized by $\Im \chi$ spectrum focused more towards higher wavelengths relative to TiO$_2$. The bound of \eqref{nbound_loss} is more closely approachable for materials with a sharp absorption peak situated as close as possible to the frequency of interest.

\section{Bianisotropic media} \label{sec:mag}

To this point, we have considered only the refractive indices of isotropic, nonmagnetic media. Although intrinsic magnetism is small at optical frequencies, the fact that patterned metamaterials can exhibit sizeable effective permeabilities suggests the possibility for magnetic response to elevate a metamaterial's \emph{effective} refractive index beyond our bounds. More generally, natural and especially artificial materials can demonstrate extreme anisotropy and magneto-electric coupling (chirality) in their response. In this section we consider the most general class of bianisotropic materials, however, and we outline a broad set of conditions under which the refractive-index bounds for such materials are identical to those of \eqreftwo{nbound}{nbound_high} discussed above.

One possibility is to simply use the refractive-index Kramers--Kronig relation and sum rule, as described in \secref{nKK}. The refractive index itself allows for magnetism and anisotropy, and thus certainly the bound of \eqref{nkkbound} would be valid for each diagonal component of the anisotropic material. Yet it turns out to again be too loose, as we will discuss below. Another possibility would be to consider a magnetic-susceptibility Kramers-Kronig relation and sum rule, in analogy to the electric-susceptibility versions of \eqref{kk} and \eqref{fsum}. However, there is no known sum rule on the imaginary part of the magnetic susceptibility. This relates to a deep and fundamental asymmetry between magnetic and electric properties of materials, and to the fact that permeability itself is not a well-defined quantity at very high frequencies~\cite{Landau2013}.

Instead, we exploit the fact that, in macroscopic electrodynamics, any bianisotropic linear material can always be described, equivalently, by purely electric spatially dispersive constitutive relations, as recognized in Refs.~\cite{Agranovich2004,Silveirinha2007,Agranovich2013}. We consider an arbitrary linear, local, bianisotropic medium, with constitutive relation
\begin{align}
\begin{pmatrix}
        \Dv \\
        \Bv
    \end{pmatrix} = 
  \begin{pmatrix}
        \epsten  & \frac{1}{c} \xiten \\
        \frac{1}{c} \zetaten & \muten
    \end{pmatrix}  
    \begin{pmatrix}
        \Ev \\
        \Hv
    \end{pmatrix},
    \label{eq:biani} 
\end{align}
where $\epsten$ is permittivity, $\muten$ is permeability, $\xiten$ and $\zetaten$ are magneto-electric coupling tensors, and $c$ is the speed of light. There is not a unique mapping from the microscopic Maxwell equations in a material (in terms of induced currents in free space) to a macroscopic description in terms of constitutive parameters, as in \eqref{biani}; in particular, it has been shown that a local, bianisotropic medium is equivalent to a \emph{nonlocal}, anisotropic, nonmagnetic medium. The nonlocality manifests through a spatially dispersive permittivity that is a function of wavevector $\kv$, with the nonlocal effective permittivity given by
\begin{align}
    \epsnl =& \ \epsten - \xiten \cdot \muten^{-1} \cdot \zetaten + \left( \xiten \cdot \muten^{-1} \times \frac{\mathbf{k}}{k_0} - \frac{\mathbf{k}}{k_0} \times \muten^{-1} \cdot \zetaten \right) \nonumber \\
             & + \frac{\mathbf{k}}{k_0} \times \left( \muten^{-1} - \overline{\overline{\mathbf{I}}} \right) \times \frac{\mathbf{k}}{k_0},
    \label{eq:epsnl} 
\end{align}
where $k_0 = \omega/c$ is the wavenumber in the host medium (taken to be vacuum). In general, \eqref{epsnl} is anisotropic even for isotropic permittivity and/or permeability, due to the wavevector dependence.  In this case, we can utilize the fact that Kramers--Kronig relations and the $f$-sum rule are valid for each diagonal component and each individual wavevector of a spatially dispersive, anisotropic medium~\cite{Altarelli1972,Agranovich2013} (cf. {\SM}). We can then represent the nonlocal susceptibility, $\chinl \equiv \epsnl - \overline{\overline{I}}$, where $\overline{\overline{I}}$ is the identity tensor, as a sum of lossless Drude--Lorentz oscillators, exactly analogous to \eqref{sumosc}. This is because we can always choose a polarization basis for which $\chinl$ is diagonal, since it is Hermitian in the absence of dissipation. (Note that $\chinl$ need not be diagonal for all frequencies $\omega$ and/or wavevectors $\mathbf{k}$ under the same basis. However, we only require that $\chinl$ is diagonalizable at a given frequency and wavevector.)

The refractive index of an anisotropic medium is itself anisotropic, and depends also on the polarization of the electromagnetic field. Consider a propagating plane wave with wavevector $\kv = \frac{n\omega}{c}\mathbf{\hat{s}}$. The square of the bianisotropic refractive index, $n_{\rm bianiso}$, experienced by that plane wave is one of two non-trivial solutions of the eigenproblem (cf. Ref.~\cite{Born2013}, also see {\SM}),
\begin{align}
\epsnl \mathbf{e_0} = n_{\textrm{bianiso}}^2 B \mathbf{e_0},
    \label{eq:geneig} 
\end{align}
where $B = \id - \mathbf{\hat{s}}\mathbf{\hat{s}}^T$ and $\mathbf{e}_0$ is the corresponding eigenvector that physically represents an eigen-polarization. For any material described by a positive- or negative-semidefinite $\epsnl$, the square of the refractive index in \eqref{geneig} is bounded by the largest eigenvalue of $\epsnl$ (we defer the discussion of indefinite $\epsnl$ to the end of this section). Choosing a polarization basis for which $\epsnl$ is diagonal, the largest eigenvalue of $\epsnl$ is its largest diagonal component. The magnitude of the diagonal components is bounded by their KK relations and sum rules, which individually degenerate to the isotropic bounds $n_{\rm max,iso}$. (This sequence of steps is mathematically proven in the SM.) Hence, the bianisotropic refractive index is bounded above by the isotropic-material bound:
\begin{align}
    n_{\textrm{bianiso}} \leq n_{\textrm{max},\textrm{iso}}. \label{eq:max_eig}
\end{align}
\Eqref{max_eig} says that, no matter how one designs bianisotropic media, its maximum attainable refractive index, for any propagation direction and polarization, can never surpass that of isotropic, electric media, as long as $\epsnl$ is positive- or negative-semidefinite. We can intuitively explain why magnetism, chirality, and other bianistropic response cannot help increase the refractive index. Instead of viewing them as distinct phenomena, it is helpful to view them as resulting from the same underlying matter, which can be distributed in different ways to create different induced currents under the action of an applied electromagnetic field. For example, one can tailor the spatial dispersion of permittivity to obtain strong magnetic dipole moments, resulting in effective permeability, or alternatively, create strong chiral response, while the number of available electrons is always the same. Independent of the resulting bianisotropic response, they can all be described by the effective, nonlocal permittivity of \eqref{epsnl} (with varying degrees of spatial dispersion), which is still subject to our upper bound based on the total available electron density. Carrying over our bound techniques employed in \secref{sf}, the maximal refractive index for such $\epsnl$ is therefore identical to \eqref{nbound} with dispersion corresponding to the maximum principal component of $\epsnl$. We show in the SM that most bianisotropic media are captured by positive-definite $\epsnl$ and also identify particular conditions (for example, magnetic materials with permeability greater than unity) under which $\epsnl$ must be positive definite. Thus, our refractive-index bound is applicable to generic bianisotropic media that describe a wide range of metamaterials. This is a powerful result suggesting that, no matter how one designs metamaterials to include magnetic, chiral, or other bianisotropic response, the tradeoff between refractive index and dispersion is inevitable.

The class of materials that have indefinite material tensors is exactly the class of \emph{hyperbolic (meta)materials}~\cite{Poddubny2013,Ferrari2015}. In such materials, the bound of \eqref{max_eig} does not apply, and in fact there is no bound that can be derived. Mathematically, this makes sense: the indefinite nature of such materials leads to hyperbolic dispersion curves that can have arbitrarily large wavenumbers at finite frequencies, and consequently refractive indices approaching infinity. Yet, physically, such waves are difficult to access as they are well outside the free-space light cone. Considering more realistic material models, based on microscopic and quantum-plasmonic considerations, this behavior is regularized by the introduction of (i) additional nonlocal effects, e.g., hydrodynamic nonlocalities, which result in a large-wavevector cutoff in the material response~\cite{Yan2012} and (ii) dissipation (e.g., Landau damping for large wavevectors). An interesting pathway forward would be to use computational optimization, e.g. ``inverse design''~\cite{Jensen2011,Miller2012,inversedesign_review,Phan2019,Chung2020}, to identify in-coupling and out-coupling structures that enable access to the high-index modes without reducing the index of the modes themselves.

Another case in which our bound does not hold is for gyrotropic plasmonic materials, the simplest example being a magnetized Drude plasma. Any conducting material has a pole at zero frequency that contributes an additional term in the KK relation for $\Im \chi$, but in gyrotropic plasmonic materials the zero-frequency pole can modify the KK relation for $\Re \chi$, altering \eqref{kk} and the subsequent analysis~\cite{Abdelrahman2020}. Due to this additional term in the KK relation for $\Re \chi$, one can attain very large values of permittivity, and hence refractive index, below the cyclotron resonance frequency with low loss and zero dispersion far away from resonance. Yet, such response only occurs below the cyclotron resonance frequency, which is typically much smaller than optical frequencies of interest for technologically available magnetic fields. 



\section{Designing high-index composites} \label{sec:highindex}
\begin{figure*} [t!]
    \includegraphics[width=0.8\linewidth]{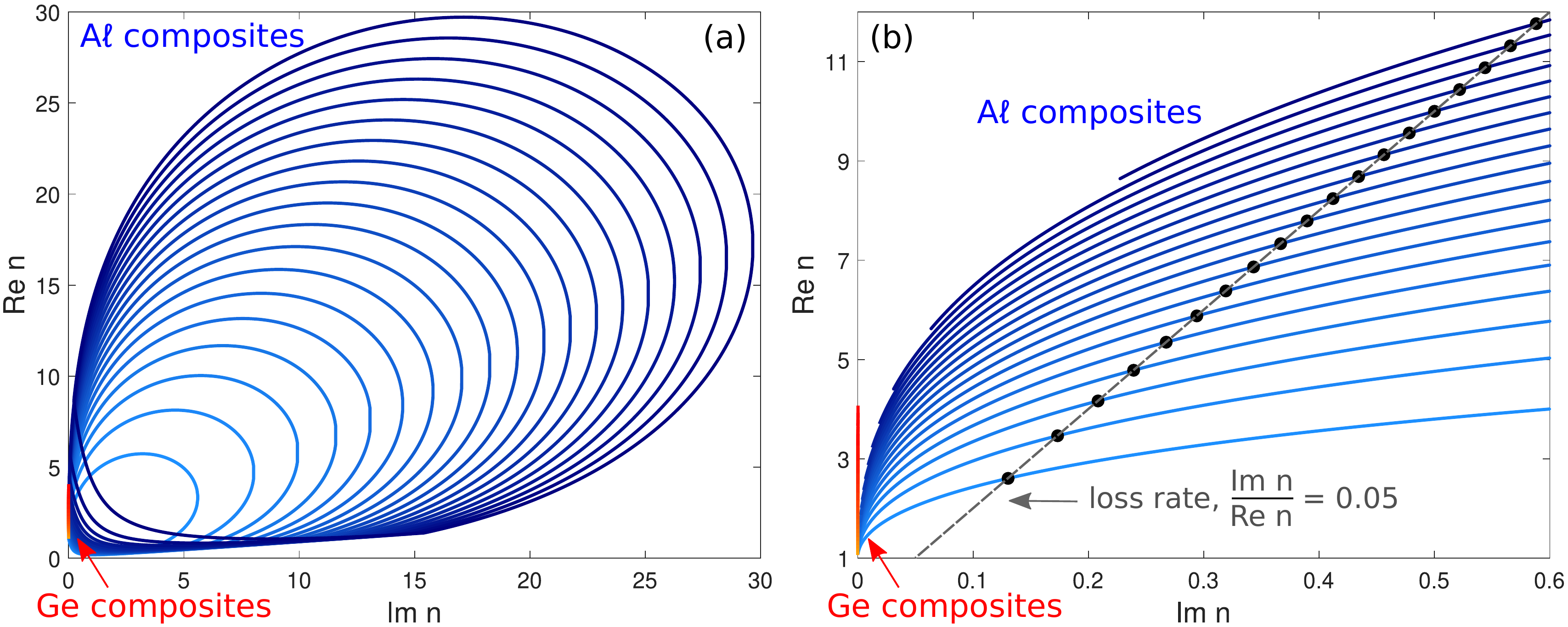} 
    \caption{The Bergman--Milton bounds~\cite{Bergman1980,Milton1980}, recently strengthened~\cite{Kern2020}, identify the feasible effective material properties of isotropic composite materials (metamaterials). (a) Feasible regions for composites of germanium (red) versus aluminum (blue) at \SI{1550}{nm} wavelength; in the latter case, each enclosed region represents a different fill fraction of aluminum relative to air. The large, negative susceptibility of aluminum enables strikingly large regions of high index, albeit also with nonzero losses. (b) The low-loss portion of the feasible regions}
    \label{fig:circlebound} 
\end{figure*}
In the previous sections we showed that for low to moderate dispersion values, natural materials already nearly saturate the fundamental bounds to refractive index. The high-dispersion, high-index part of the fundamental-limit curve has no comparison points, however, as there are no materials that exhibit high dispersion in transparency windows at optical frequencies, and hence no materials exhibit the high refractive indices our bounds suggest should be possible. In fact, renormalization-group principles~\cite{Andreoli2021} have been used to identify the maximum refractive index in ensembles of atoms, yielding a value 1.7 that is close to those of real materials. Hence, an important open question is whether it is possible to engineer high refractive index, even allowing for high levels of dispersion?

Here, we show that composite materials can indeed exhibit significantly elevated refractive indices over their natural-material counterparts. Key to the designs is the use of metals and negative-permittivity materials, whose large susceptibilities unlock large positive refractive indices when patterned correctly. We find that with typical metals such as silver and aluminum, it should be possible to reach refractive indices larger than 10, with small losses, at the telecommunications wavelength \SI{1.55}{\um}. The lossiness of the metals is the only factor preventing them from reaching even larger values; if it becomes possible to synthesize the ``elusive lossless metal''~\cite{Khurgin2010}, with vanishingly small loss, then properly designed composites can exhibit refractive indices of 100 and beyond.
\begin{figure*} [tbh]
    \includegraphics[width=0.9\linewidth]{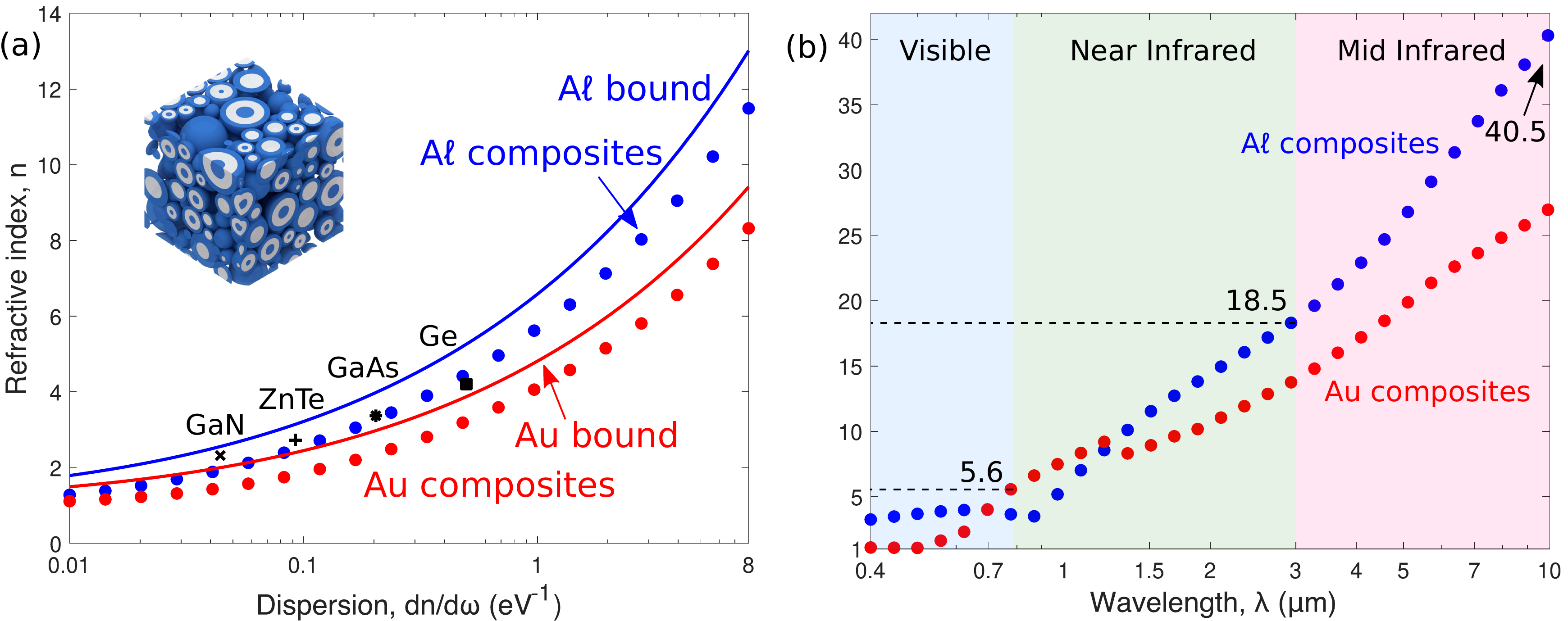}
    \caption{Composites can achieve high refractive indices, at high levels of dispersion, as predicted by our bounds. (a) At \SI{1550}{nm} wavelength, typical high-index dielectrics such as GaAs and Ge have refractive indices approaching 4. By contrast, assemblages of doubly coated spheres (inset) of gold and aluminum can be designed to achieve low-loss, effective refractive indices above 8 and approaching 12, respectively. Moreover, these composites quite closely approach our bounds (solid lines), suggesting that they are tight or nearly so. (b) Maximum low-loss refractive index of gold and aluminum composites as a function of wavelength. Much higher refractive indices are possible at longer wavelengths, as predicted by our bounds.}
    \label{fig:composite} 
\end{figure*} 
\begin{figure} [tbh]
    \includegraphics[width=0.9\linewidth]{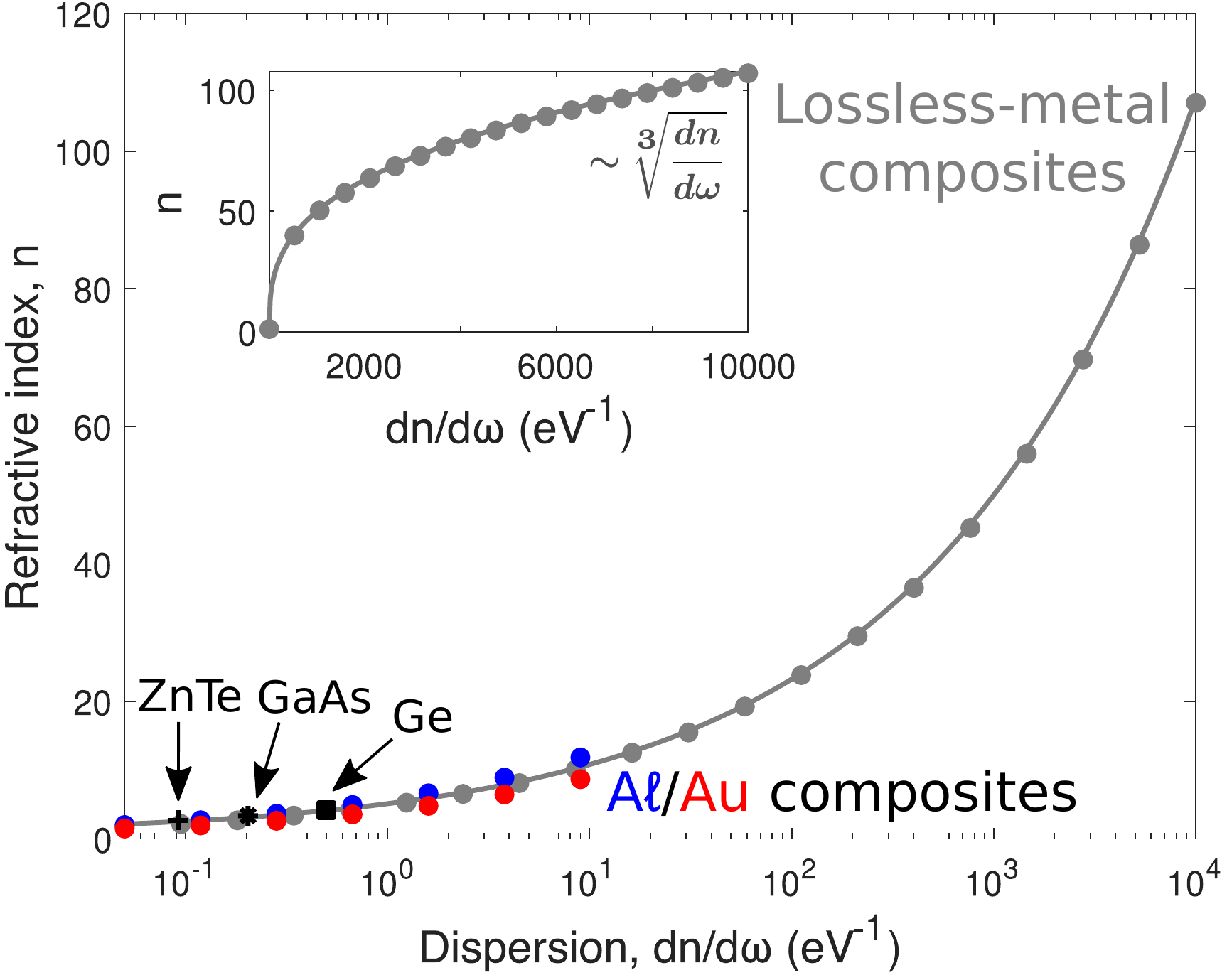}
    \caption{Lower-loss metals would enable even more dramatic enhancements of refractive index. Composites with a nearly lossless metal can be designed to achieve refractive indices larger than 100 at \SI{1550}{nm} wavelength. These composites (circle markers) exactly achieve our bounds (solid line), and require enormous dispersion values to do so, thanks to the cube-root scaling indicated in the inset.}
    \label{fig:lossless} 
\end{figure}

The theory of composite materials and the effective material properties that can be achieved has been developed over many decades~\cite{Milton2002,Milton2016}. Composite materials, or metamaterials, comprise multiple materials mixed at highly subwavelength length scales that show effective properties different from those of their underlying constituents. They offer a promising potential route, then, to achieving higher refractive indices through mixing than are possible in natural materials themselves. Bounds, or fundamental limits, to the possible refractive index of an isotropic composite have been known since the pioneering work of Bergman and Milton~\cite{Bergman1980,Milton1980,Milton1981,Milton1981a,Bergman1981} (and even earlier for lossless materials~\cite{Hashin1962}), and were recently updated and tightened~\cite{Kern2020}. Bounds are identified as a function of the fill fraction of one of the two (or more) materials. For composite of two materials, the bounds comprise two intersecting arcs in the complex permittivity plane. The analytical expressions for the bounds are given in Eqs.~(7,79) of \citeasnoun{Kern2020}, which we do not repeat here due to their modest complexity. 
 
In \figref{circlebound}, we demonstrate what is possible according to the updated Bergman--Milton bounds. At \SI{1550}{nm} wavelength, we consider two classes of composite, one comprising a higher index dielectric material, germanium, with a low-index material taken to be air, and the second comprising a metallic material, aluminum, with the same air partner. The Ge-based composite exhibits only small variations in its possible refractive index, the red line, occupying the range between 1 (air) and 4.2 (Ge). By contrast, composites with aluminum can exhibit far greater variability, and potentially much larger real parts of their refractive index. The increasingly large regions occupied by the blue arcs represent the bound regions with increasing fill fractions of the aluminum. Of course, one cannot simply choose the highest real refractive index: most of those points are accompanied by tremendously large loss as well. Part (b) of \figref{circlebound} zooms in on the lower left-hand side of the complex-$n$ plane, where the imaginary parts are sufficiently small that the materials can be considered as nearly lossless. In that region, one can see that there are still sizable possible refractive indices. The largest loss rate can be defined as a ratio of the imaginary part of $n$ to its real part. The real part determines the length over which a $2\pi$ phase accumulation can be achieved, while the imaginary part determines the absorption length, and the key criteria would typically be a large ratio of the two lengths. The black line in \figref{circlebound}(b) represents a loss-rate ratio of $\Im n / \Re n = 0.05$. One can see that refractive indices beyond 11 are achievable with an Al-based composite.

It is important to emphasize that the refractive indices shown in \figref{circlebound}(b) are indeed achievable. All of the low-loss bounds shown there, and below, arise from the circular arc that is known to be achievable by assemblages of doubly coated spheres~\cite{Kern2020}. The inset of \figref{composite}(a) schematically shows such an assemblage, comprising densely packed doubly coated spheres that fill all space (cf. Sec.~(7.2) of \citeasnoun{Milton2002}). \Figref{composite}(a) uses circular markers to indicate the largest refractive indices that are possible, as a function of their dispersion values, for doubly-coated-sphere assemblages of aluminum and gold. (Silver is very similar to gold in its possible refractive-index values, due to their similar electron densities.)  Accompanying the markers are solid lines that indicate the electron-density-based refractive-index bounds of \eqref{nbound}. One can see that the composites track quite closely with the bounds. Also included are markers for some of the highest-index natural materials, GaN, ZnTe, and GaAs, clearly showing the dramatic extent to which metal-based composites can improve on their natural dielectric counterparts. The figure does not go past dispersion values of \SI{8}{eV^{-1}}, however, as the losses of the composites grow too large in the designs for higher dispersion values. In \figref{composite}(b), we map out the largest refractive indices as a function of wavelength that are possible with low-loss composites, with loss rates, as defined above, no larger than 0.05. With such composites, refractive indices larger than 5, 18, and 40 are possible in the visible, near-infrared, and mid-infrared frequency ranges, respectively. Each would represent a record high in its respective frequency range.

The large indices of the Al- and Au-based composites can be increased even further with lower-loss materials. To test the limits of what is possible, in \figref{lossless} we consider a composite with a lossless Drude metal with plasma frequency of \SI{10}{eV} (corresponding to an electron density of 0.7$\times$10$^{23}$$\,$cm$^{-3}$). The updated BM bound, achieved by the doubly-coated-sphere assemblages, can now exhibit phenomenally large refractive indices, even surpassing 100 in the infrared. As required by our bounds, such refractive indices are accompanied by phenomenally large dispersion values, and the inset shows the slow cube-root increase of refractive index with dispersion for these composites. Our bound of \eqref{nbound}, applied to the Drude material, now lies along the curve for the composites, showing that the composites can saturate our bounds (and, consequently, that our bounds are tight and cannot be further improved.) There is significant interest in engineering lossless metals~\cite{Khurgin2010,Boltasseva2011}; if it can be done, we have shown that refractive indices above 100 would be achievable at optical frequencies. 

\section{Summary and extensions} \label{sec:conclusion}
We have established the maximal refractive index valid for arbitrary passive, linear media, given constraints on dispersion or bandwidth. Starting from Kramers--Kronig relations and the $f$-sum rule that all causal media have to obey, we have obtained a general representation of susceptibility. We have employed linear-programming techniques to demonstrate that the optimal solution is a single Drude--Lorentz oscillator with infinitesimal loss rate, which gave simple, analytic bounds on refractive index. Based on a similar approach, we have obtained bounds on high-index optical glasses and refractive index averaged over arbitrary bandwidth. We have also generalized our bounds to any bianisotropic media described by a positive- or negative-semidefinite effective permittivity $\epsnl$, rendering our bounds more general than initially expected (i.e., the maximal refractive indices obtained in \secref{sf} and \secref{bw} also describe materials incorporating magnetic, chiral, and other bianisotropic response). We have also designed low-loss metal-based composites with refractive indices exceeding those of best performing natural materials by a factor of two or more in the high-dispersion regime.

The approach developed herein can be extended to address a variety of related questions. For example, one can allow for gain media, which can still be described by a sum of Drude--Lorentz oscillators with infinitesimal loss rates (see \eqref{sumosc}). However, the oscillator strengths in this scenario need not be positive, leading to different optimal linear-programming solutions depending on the exact objective and constraints. In the case of gain media, stability considerations become crucial, as a high bulk refractive index, or any other bulk property, may be irrelevant, if the resulting structure exhibits an unstable response with unbounded temporal oscillations~\cite{Nistad2008}. Besides, while we have considered optical frequencies in this paper, the bounds established here can be used to compare state-of-the-art dielectrics at microwave and other frequencies of interest. One may also be interested in metrics other than refractive index. A key metric in the context of waveguides and optical fibers is group velocity dispersion~\cite{Okamoto2006}, which can be seamlessly incorporated into our framework. 

Another metric closely related to refractive index is the group index, which measures the reduction in $\emph{group}$ velocity of electromagnetic waves in a medium. However, unlike refractive index, the group index can reach values up to $60$ even in the near-IR, and much higher elsewhere~\cite{Baba2008}. This is because group index $n_g$, by definition, increases with dispersion:
\begin{align}
n_g = \frac{{\rm d} (n\omega)}{{\rm d} \omega} = n + \omega \frac{{\rm d} n}{{\rm d} \omega}. \label{eq:group_index}
\end{align}
Since the first term in \eqref{group_index} is just the refractive index, which is often of the order of unity, the second term, scaling with dispersion, is usually the dominant term for very large values of group index. That being said, we show in the {\SM} that bounds on group index averaged over arbitrary nonzero bandwidth can be obtained based on our refractive-index bound.

One can also explore negative (anomalous) dispersion, which typically occurs around resonances where losses are sizeable. To do so, one might construct other representations (such as B-splines~\cite{Ivanenko2019}) that are more suited to describe regions of near-zero or negative dispersion. 

An intriguing alternative extension is to nonlinear material properties. There are known Kramers--Kronig relations for nonlinear susceptibilities~\cite{Boyd2003}, yet their sum rules~\cite{Lucarini2005} are more complex than those of linear susceptibilities. If the sum rules can be simplified, or even just bounded, then it should be possible to identify bounds on nonlinear susceptibilities.

Another avenue that can potentially prove fruitful is to better understand the key characteristics of materials that determine refractive index. While the maximum allowable dispersion sets a limit on refractive index, are there more fundamental, physical quantities at play behind the scene? In the {\SM}, we identify a characteristic trait of high-index materials: a combination of low molar mass and high electronegativity, to achieve large valence electron densities. Going further, it might prove fruitful to combine the insights and directions laid out here with band-structure analysis (through \textit{ab-initio} methods for example), to extract physical properties conducive to high-index materials.

Finally, direct experimental demonstrations of the high-index materials proposed in \secref{highindex} would represent record refractive indices. Techniques such as ``inverse design''~\cite{Jensen2011,Miller2012,inversedesign_review,Phan2019,Chung2020} may enable identification of structures with similarly high refractive indices in architectures more amenable to fabrication than assemblages of doubly coated spheres. Identifying such composite materials would open new possibilities in areas from metasurface optics to high-quality-factor resonators. Each of these fields could benefit even more dramatically, potentially, with the discovery or synthesis of a near-zero-loss metal, which, as we have shown, could offer refractive indices approaching 100 at optical frequencies.

\section{Acknowledgments}

We thank Christian Kern for providing the illustration of an assemblage of doubly coated spheres. We thank Jacob Khurgin, Michael Fiddy, and Richard Haglund for helpful conversations. H.S. and O.D.M. were supported by the U.S. Defense Advanced Research Projects Agency and Triton Systems. F.M. was supported by the Air Force Office of Scientific Research under Grant FA9550-19-1-0043.




%

\end{document}


\preprint{APS/123-QED}

\title{Supplementary Materials: Fundamental limits to refractive index at optical frequencies}

\author{Hyungki Shim}
\affiliation{Department of Applied Physics and Energy Sciences Institute, Yale University, New Haven, Connecticut 06511, USA}
\affiliation{Department of Physics, Yale University, New Haven, Connecticut 06511, USA}
\author{Francesco Monticone}
\affiliation{School of Electrical and Computer Engineering, Cornell University, Ithaca, New York 14853, USA}
\author{Owen D. Miller}
\email{owen.miller@yale.edu}
\affiliation{Department of Applied Physics and Energy Sciences Institute, Yale University, New Haven, Connecticut 06511, USA}

\date{\today}

\pacs{Valid PACS appear here}
\maketitle

\tableofcontents

\section{General representation of susceptibility based on discretizing Kramers--Kronig relations}  \label{sec:genrep}

In this section, we derive the general representation for susceptibility in the main text, which is valid for arbitrary linear media. Our starting point is Kramers--Kronig relations that all causal materials obey:
\begin{align}
\Re \chi(\omega) = \frac{2}{\pi}  \int_0^\infty \frac{\omega' \Im \chi(\omega')}{\omega'^2 - \omega^2} \,{\rm d}\omega',  \label{eq:kk} 
\end{align}
where the integral here is a principal-value integral (as discussed in the main text, certain plasmonic gyrotropic materials can have an extra term arising from the pole at zero frequency, but this is irrelevant at optical frequencies). Without further information on the numerator $\omega \Im \chi(\omega)$, \eqref{kk} does not tell us much about the general form of $\Re \chi$. To better represent $\Re \chi$, we use the $f$-sum rule~\cite{King1976}, which relates the imaginary part of susceptibility integrated over all frequencies to the system's plasma frequency $\omega_p$:
\begin{align}
\int_0^\infty \omega \Im \chi(\omega) \,{\rm d}\omega = \frac{\pi \omega_p^2}{2}.  \label{eq:fsum} 
\end{align}
\Eqref{fsum} implies that we are free to discretize $\omega \Im \chi(\omega)$ into arbitrary localized basis functions, as long as their sum is constrained to equal $\frac{\pi \omega_p^2}{2}$. In the limit of perfect localization, $\omega \Im \chi(\omega)$ can be written as a sum of delta functions as follows: 
\begin{align}
\omega \Im \chi(\omega) = \frac{\pi \omega_p^2}{2} \sum_{i=1}^N c_i \delta(\omega - \omega_i).  \label{eq:loc} 
\end{align}
We take out the $\frac{\pi \omega_p^2}{2}$ in \eqref{loc} so that the $f$-sum rule is now conveniently encoded in the following equation:
\begin{align}
\sum_{i=1}^N c_i = 1.  \label{eq:csum} 
\end{align}
Inserting \eqref{loc} into \eqref{kk} leads to the general representation of $\Re \chi$ as follows (note that in this representation, passive media amounts to enforcing $c_i \geq 0$ for $i=1,...,N$):
\begin{align}
\Re \chi(\omega) = \sum_{i=1}^N  \frac{c_i \omega_p^2}{\omega_i^2 - \omega^2}.  \label{eq:sumosc} 
\end{align}			 

It is straightforward to verify that \eqref{sumosc} is equivalent to a sum of Drude--Lorentz oscillators with infinitesimal loss rates. To show this, we consider both the real and imaginary part of the susceptibility obtained above:
\begin{align}
\chi(\omega) &= \sum_{i=1}^N  \frac{c_i \omega_p^2}{\omega_i^2 - \omega^2} + i \frac{\pi \omega_p^2}{2} \sum_{i=1}^N c_i \frac{\delta(\omega - \omega_i)}{\omega} \nonumber \\
&= \sum_{i=1}^N  c_i \omega_p^2 \left( \frac{1}{\omega_i^2 - \omega^2} + \frac{i \pi}{2}  \frac{\delta(\omega - \omega_i)}{\omega}  \right) \nonumber \\
&= \sum_{i=1}^N  c_i \omega_p^2 \left( \frac{1}{\omega_i^2 - \omega^2} + i \pi  \delta(\omega_i^2 - \omega^2)  \right) \nonumber \\
&= \sum_{i=1}^N  c_i \lim_{\gamma \rightarrow 0} \frac{\omega_p^2}{\omega_i^2 - \omega^2 - i \gamma \omega}, \label{eq:osc_dl}
\end{align}
where we have used the fact that $\delta(\omega_i^2 - \omega^2) = \frac{\delta(\omega_i - \omega)}{2\omega}$ (we drop the other term proportional to $\delta(\omega_i + \omega)$, since $\omega \geq 0$) in going from the second to third line, the identity $\lim_{\gamma \rightarrow 0} \frac{1}{x +i\gamma} = \frac{1}{x} -i\pi \delta(x)$ from the third to the last line, and let $\gamma \rightarrow 0$ from above. Thus, \eqref{osc_dl} explicitly shows that the general susceptibility is represented by a sum of lossless Drude--Lorentz oscillators.

\section{Proof that a single oscillator achieves maximal refractive index} \label{sec:proofosc}

Recall that our refractive-index problem is given by: 
\begin{equation}
    \begin{aligned}
        & \underset{c_i}{\text{maximize}} & &  \Re \chi(\omega) = \sum_{i=1}^N c_i \, f_i(\omega) \\
        & \text{subject to}       & & \frac{{\rm d} \Re \chi(\omega)}{{\rm d} \omega} =  \sum_{i=1}^N c_i \, g_i(\omega) \leq \chi', \\
        & & &  \sum_{i=1}^N c_i  = 1, \ c_i \geq 0, \label {eq:opt_sf}
    \end{aligned}
\end{equation} 
where $f_i(\omega) = \frac{\omega_p^2}{\omega_i^2 - \omega^2}$ and $g_i(\omega) \equiv f_i'(\omega) = 2 \omega \frac{\omega_p^2}{\left(\omega_i^2 - \omega^2 \right)^2}$.

We can show that the optimal solution to \eqref{opt_sf} is given by a single oscillator, as stated in the main text (i.e. $c_i = 1$ for only one $i$ and zero elsewhere). To this end, we first consider two oscillators and demonstrate that they can be combined into a single oscillator to give a larger susceptibility (or equivalently, refractive index). Then, we argue that this suffices to guarantee a single-oscillator optimal solution regardless of the initial oscillator number and configuration.

We start with two oscillators with oscillator frequencies $\omega_1$ and $\omega_2$ having oscillator strengths $c_1$ and $c_2$ respectively, such that $c_1 + c_2 = 1$ (we assume $\omega_2 > \omega_1$ without loss of generality). Now, we wish to show that the susceptibility arising from the two oscillators is always smaller than that for a single oscillator with the same dispersion at some frequency $\omega$ (whose oscillator frequency $\omega_0$ must lie between $\omega_1$ and $\omega_2$, and have oscillator strength of 1). For simplicity of notation, we will denote the former by $\chi_{\textrm{two}}$ and the latter by $\chi_{\textrm{one}}$, which are given by:
\begin{align}
\chi_{\textrm{two}} = c_1 f_1(\omega) + c_2 f_2(\omega) = \frac{c_1 \omega_p^2}{\omega_1^2 - \omega^2} + \frac{c_2 \omega_p^2}{\omega_2^2 - \omega^2}, \nonumber \\ 
\chi_{\textrm{one}} = f_0(\omega) = \frac{\omega_p^2}{\omega_0^2 - \omega^2}.  \label{eq:chi12} 
\end{align}
Also, all the oscillator frequencies are larger than $\omega$---otherwise, the susceptibility will become negative.

In order to compare the $\chi_{\textrm{two}}$ and $\chi_{\textrm{one}}$, it is useful to express the oscillator strengths $c_1$ and $c_2$ in terms of the oscillator frequencies. This is done by requiring that the dispersion for the two cases are equal:
\begin{align}
g_1(\omega) + g_2(\omega) = g_0(\omega),  \label{eq:dispcon} 
\end{align}	
which reduces to the following equation:
\begin{align}
\frac{c_1}{(\omega_1^2 - \omega^2)^2} +  \frac{c_2}{(\omega_2^2 - \omega^2)^2} =  \frac{1}{(\omega_0^2 - \omega^2)^2}.  \label{eq:dispmatch} 
\end{align}	
Combining \eqref{dispmatch} with the constraint that $c_1 + c_2 = 1$, we can write $c_1$ and $c_2$ as:
\begin{align}
c_1 = \frac{(\omega_1^2 - \omega^2)^2}{(\omega_0^2 - \omega^2)^2} \left[ \frac{(\omega_0^2 - \omega^2)^2 - (\omega_2^2 - \omega^2)^2}{(\omega_1^2 - \omega^2)^2 - (\omega_2^2 - \omega^2)^2} \right], \nonumber \\
c_2 = \frac{(\omega_2^2 - \omega^2)^2}{(\omega_0^2 - \omega^2)^2} \left[ \frac{(\omega_1^2 - \omega^2)^2 - (\omega_0^2 - \omega^2)^2}{(\omega_1^2 - \omega^2)^2 - (\omega_2^2 - \omega^2)^2} \right].  \label{eq:c1c2} 
\end{align}	
\Eqref{c1c2} gives the required value of $\omega_0$, for arbitrary choices of coefficients $c_1$ and $c_2$ that sum up to 1, so that the single-oscillator dispersion is equal to that for two oscillators. We can substitute \eqref{c1c2} in \eqref{chi12} to obtain an explicit expression for the ratio $\chi_{\textrm{two}} / \chi_{\textrm{one}}$, which determines whether a single oscillator has higher susceptibility than that for two oscillators:
\begin{align}
\frac{\chi_{\textrm{two}}}{\chi_{\textrm{one}}} = \frac{ (\omega_2^2 - \omega^2) \left[ (\omega_1^2 - \omega^2)^2 - (\omega_0^2 - \omega^2)^2 \right] + (\omega_1^2 - \omega^2) \left[ (\omega_0^2 - \omega^2)^2 - (\omega_2^2 - \omega^2)^2 \right] }{(\omega_0^2 - \omega^2) \left[ (\omega_1^2 - \omega^2)^2 - (\omega_2^2 - \omega^2)^2 \right]}.  \label{eq:chiratio} 
\end{align}	
To simplify notation, let us define the following symbols: $\alpha \equiv \omega_1^2 - \omega^2$, $\beta \equiv \omega_2^2 - \omega^2$, and $\gamma \equiv \omega_0^2 - \omega^2$. Then, \eqref{chiratio} can be written as:
\begin{align}
\frac{\chi_{\textrm{two}}}{\chi_{\textrm{one}}} &= \frac{\beta (\alpha^2 - \gamma^2) + \alpha (\gamma^2 - \beta^2))}{\gamma (\alpha^2 - \beta^2)}. \nonumber \\
&= \frac{\gamma^2 + \alpha\beta}{\gamma (\alpha + \beta)} < 1.  \label{eq:chiratio2} 
\end{align}	
\Eqref{chiratio2} establishes that a single oscillator always has higher susceptibility (or equivalently, refractive index) than two oscillators. To see this, we can use the fact that $(\gamma - \alpha)(\gamma-\beta)<0$ (since $\omega_1 < \omega_0 < \omega_2$ is a necessary condition for \eqreftwo{dispcon}{dispmatch} to hold, and this implies that $\alpha < \gamma < \beta$). Expanding this expression, we have that $\gamma^2 + \alpha \beta < \gamma(\alpha + \beta)$, which is precisely the condition needed for the inequality in the last line of \eqref{chiratio2}.

We can easily generalize this result to $N$ ($>2$) oscillators by picking any two oscillators, replacing them by one oscillator (so now there are $N-1$ oscillators), and iterating the whole process until only one oscillator remains, which will necessarily have the largest susceptibility (with, of course, the same dispersion as the initial value). Thanks to the linearity of the problem, the optimal solution for our refractive-index problem in \eqref{opt_sf} is a single oscillator, no matter how many oscillators we allow.

\section{Alternative proof based on Lagrangian duality} \label{sec:lag}

In this section, we use Lagrangian duality to show how the cardinality of the optimal solution---number of distinct oscillators in our case---is related to the nature of objective and constraint functions. For our refractive-index problem, we then explicitly derive general conditions under which the optimal solution is a single oscillator. 

First, let us set up the (primal) problem to consider, where the primal variable, objective, and constraint functions are denoted by $\mathbf{c}$, $\mathbf{f}$, $\mathbf{g}_i$ respectively. All of them are $N$-dimensional vectors, and $i=1,...,M$, where $N$ and $M$ denote the dimensionality and the number of inequality constraints respectively. We wish to solve the following problem:
\begin{equation}
    \begin{aligned}
        & \underset{\mathbf{c}}{\text{maximize}} & & \mathbf{f}^T \mathbf{c} \\
        & \text{subject to}       & &  \mathbf{g}_i^T \mathbf{c} \leq b_i \quad i=1,...,M, \\
        & & &  \mathbf{1}^T \mathbf{c}  = 1, \\
        & & & c_i \geq 0 \quad i=1,...,N,  \label {eq:primal}
    \end{aligned}
\end{equation} 
where $\mathbf{1}$ is a $N \times 1$ vector of ones, and we have separated out the equality constraint as it is particularly relevant for our refractive-index problem (i.e., it encodes the $f$-sum rule). The dual problem of \eqref{primal} is given as follows, where $\mathbf{y}$ is a dual variable of dimension $M+1$~\cite{Vanderbei2015}:
\begin{equation}
    \begin{aligned}
        & \underset{\mathbf{c}}{\text{minimize}} & & (\mathbf{b}^T,1) \cdot \mathbf{y}  \\
        & \text{subject to}       & &  (\mathbf{g}_i^T, 1) \cdot \mathbf{y} \leq f_i \quad i=1,...,N, \\
        & & & y_i \geq 0 \quad i=1,...,M, \ y_{M+1} \in \mathbb{R}.  \label {eq:dual}
    \end{aligned}
\end{equation} 
A useful property of linear programming is strong duality~\cite{Vanderbei2015}, which means that the maximum value of objective in \eqref{primal} is equal to the minimum value of objective in \eqref{dual}. Thus, \eqref{primal} and \eqref{dual} are equivalent as far as the optimal objective value is concerned (the optimal solution may not be unique, but this does not matter---the dual solution at least provides $\emph{an}$ optimal solution, which suffices for our purposes). Notice that the primal constraints map to dual variables and vice versa---\eqref{primal} has $N$ variables and $M+1$ constraints, whereas \eqref{dual} has $M+1$ variables and $N$ constraints (excluding the non-negativity constraint). For each nonzero component of $\mathbf{c}$ ($c_i \neq 0$), the corresponding $g_i$ becomes a constraint in dual space. Otherwise, if $c_i = 0$, one can drop $g_i$ from the primal problem without affecting the optimal value (or, the $g_i$ constraint is inactive in dual space). We thus arrive at the following conclusion: \emph{the cardinality of optimal solution is equal to the number of active dual constraints}. 

In general, determining how many dual constraints are active is not straightforward, and depends on the nature of the objective and constraint functions $\mathbf{f},\mathbf{g}$. After all, the dual problem itself is a linear programming, which is typically as difficult to solve as the primal problem. That being said, one can derive explicit conditions on the cardinality of optimal solution when there are only 2 primal constraints (i.e., $M=1$), or equivalently, the dual variable $\mathbf{y}$ is two-dimensional. This is the case for the refractive-index problem considered in the main text, and so we take $M=1$ in what follows.

\subsection{Two primal constraints (scalar $g$)}
The optimal solution of \eqref{primal} with $M=1$ has cardinality of at most two, and generally occurs at a vertex---intersection of two lines---in dual space. In certain cases, the optimal solution can lie on a line or at the intersection with the axis along $y_2$ (since $y_1 \geq 0$ and $y_2 \in \mathbb{R}$). We can render this notion more precise: the optimal solution has cardinality of 2, and cardinality of 1 under two conditions described below (assuming $f_i,g_i >0$ for simplicity):

\subsubsection{Trivial case}
If the largest component of the objective vector $\mathbf{f}$ has a corresponding component of $\mathbf{g}$ not exceeding the constraint cutoff $b$ (i.e., if there exists $k$ such that $f_k \equiv \max{f}$ has $g_k \leq b$), the optimal solution in dual space is given by the intersection of the $g_k$ constraint with the $y_2$-axis. Thus, there is only one active dual constraint, or translated to the primal problem, the optimal solution has cardinality of one. In hindsight, this is a rather trivial case since the $g_k$ constraint is not binding, and one can simply choose $c_k =1$ along the component $k$ with largest value of $\mathbf{f}$.

\subsubsection{Physically relevant case}
For the case where $g_n = b$ for some $n \in [1,N]$, and $g_i >b$ if $f_i > f_n$ for all $i \neq n$ (which is certainly true for our refractive-index problem, where $f_i = \frac{\omega_p^2}{\omega_i^2 - \omega^2}$ and $g_i= 2 \omega \frac{\omega_p^2}{\left(\omega_i^2 - \omega^2 \right)^2}$ for a given frequency $\omega$ and plasma frequency $\omega_p$), one can derive precise, mathematical conditions for which the cardinality of optimal solution is one. Graphically, this occurs when $f_n$ is greater than, or equal to, all other vertices formed by intersection of any two constraints in dual space. If this holds, the optimal solution of \eqref{primal} is given by $c_n=1$ (in dual space, the optimal solution is a line given by the $g_n$ constraint), with the optimal value of $f_n$ with the active constraint $g_n=b$. 

Given two lines (dual constraints) of the form $(g_i, 1) \cdot \mathbf{y} = f_i$ and $(g_j, 1) \cdot \mathbf{y} = f_j$, one can solve for the intersection point $\mathbf{y}^{*} \equiv \left( \frac{f_j-f_i}{g_j-g_i}, f_i - g_i \frac{f_j-f_i}{g_j-g_i} \right)^T$. Thus, the following condition ensures that all other intersection points do not lie above $(g_n, 1) \cdot \mathbf{y} = f_n$:
\begin{align}
(g_n, 1) \cdot \mathbf{y}^{*} \leq f_n. \label{eq:vertices}
\end{align}
Substituting the expression for $\mathbf{y}^{*}$ into \eqref{vertices}, we can obtain a precise condition under which the optimal solution has cardinality of one (where we assume there exists an index $n$ such that $g_n =b$):
\begin{align}
 \frac{f_j-f_i}{g_j-g_i} \leq  \frac{f_n-f_i}{g_n-g_i} \quad \forall i,j \neq n, \quad \textrm{such that} \quad f_i < f_n < f_j, \ g_i < g_n < g_j . \label{eq:cardone}
\end{align} 
Heuristically speaking, \eqref{cardone} shows that if the cost is greater than the benefit of increasing the objective, then it is better to have a cardinality-one solution $c_n=1$. It is straightforward to verify that \eqref{cardone} holds for our refractive-index problem with dispersion constraint (as well as other related problems, see \eqref{opt_gen} for e.g.), explaining why the single-oscillator solution is optimal. For the dispersion constraint, one has $f_i = \frac{\omega_p^2}{\omega_i^2 - \omega^2}$ and $g_i= 2 \omega \frac{\omega_p^2}{\left(\omega_i^2 - \omega^2 \right)^2}$, which satisfies \eqref{cardone}. Contrary to the derivation laid out in \secref{proofosc}, the proof established here holds more generally to a wide range of linear-programming problems, independent of the specifics of the problem, whose optimal solution is guaranteed to have cardinality of one as long as \eqref{cardone} is satisfied.

\section{Derivation of refractive-index bound from a single-oscillator solution}

The optimal solution of the refractive-index problem in \eqref{opt_sf} is described by a single, lossless Drude--Lorentz oscillator. Explicitly writing the susceptibility,
\begin{align}
\Re \chi_{\textrm{opt}}(\omega) = \frac{\omega_p^2}{\omega_0^2 - \omega^2},	\label{eq:sfchi}
\end{align}
where the oscillator frequency $\omega_0$ is set by the dispersion constraint. Since higher dispersion leads to higher susceptibility, the dispersion will equal $\chi'$, its maximum allowed value (i.e. the constraint is active):
\begin{align}
\frac{{\rm d} \Re \chi_{\textrm{opt}}(\omega)}{{\rm d} \omega} = \frac{2\omega_p^2 \omega}{(\omega_0^2 - \omega^2)^2} = \chi'.	\label{eq:sfdisp}
\end{align}
Substituting \eqref{sfdisp} into \eqref{sfchi}, we can obtain the optimal profile in terms of (maximum allowable) dispersion $\chi'$ and plasma frequency $\omega_p$ (we drop the $\Re$ symbol since $\chi_{\textrm{opt}}$ is manifestly real):
\begin{align}
\chi_{\textrm{opt}}(\omega) = \frac{\omega_p}{\sqrt{2}} \sqrt{\frac{\chi'}{\omega}}.	\label{eq:sfchi_opt}
\end{align}
Since it is more common to express dispersion in terms of $n' \equiv \frac{{\rm d} n}{{\rm d} \omega}$, we use the relation $\chi' = 2 n n'$ to replace $\chi'$ with $n'$. Then, the refractive index can be bounded as follows:
\begin{align}
n = \sqrt{1 + \chi} \leq \left(1 + \omega_p \sqrt{\frac{n n'}{\omega}} \ \right)^{1/2}.	\label{eq:n_opt}
\end{align}
Rearranging \eqref{n_opt}, we obtain the following bound on refractive index:
\begin{align}
\frac{(n^2-1)^2}{n} \leq \frac{\omega_p^2 n'}{\omega},  \label{eq:nbound}
\end{align}
which further simplifies in the limit of large refractive index (i.e. $n^2 \gg 1$):
\begin{align}
n \leq \left( \frac{\omega_p^2 n'}{\omega} \right)^{1/3}.  \label{eq:nbound_high}
\end{align}


\section{Refractive-index problem over nonzero bandwidth}

So far, we have maximized refractive index at a single frequency. We can also consider maximizing refractive index (or susceptibility) over some spectrum $[\omega_a, \omega_b]$, characterized by a nonzero frequency bandwidth $\Delta \omega \equiv \omega_b - \omega_a$ with (maximum allowable) bandwidth-averaged dispersion $\overline{\chi'}$:
\begin{equation}
    \begin{aligned}
        & \underset{c_i}{\text{maximize}} & & \frac{1}{\Delta \omega}\int_{\omega_a}^{\omega_b} \Re \chi(\omega) \,{\rm d}\omega \\
        & \text{subject to}       & & \frac{1}{\Delta \omega}\int_{\omega_a}^{\omega_b} \frac{{\rm d} \Re \chi(\omega)}{{\rm d} \omega} \,{\rm d}\omega \leq \overline{\chi'},  \label {eq:opt_gen}
    \end{aligned}
\end{equation}
where, although not explicitly written above, the material is passive and satisfies the $f$-sum rule. Analogous to the single-frequency case, the optimal solution for \eqref{opt_gen} is a single, lossless Drude--Lorentz oscillator (as proved in \secref{lag}). Thus, we can explicitly write down the optimal bandwidth-averaged susceptibility and dispersion as follows: 
\begin{align}
\frac{1}{\Delta \omega}  \int_{\omega_a}^{\omega_b} \Re \chi_{\textrm{opt}}(\omega) \,{\rm d}\omega &= \frac{\omega_p^2}{\Delta \omega} \int_{\omega_a}^{\omega_b} \frac{1}{\omega_0^2 - \omega^2} \,{\rm d}\omega \nonumber \\
&= \frac{\omega_p^2}{\Delta \omega} \frac{1}{\omega_0} \left[ \tanh^{-1}\left(\frac{\omega_b}{\omega_0}\right) - \tanh^{-1}\left(\frac{\omega_a}{\omega_0}\right) \right], \label{eq:avgsus}
\end{align}
\begin{align}
\frac{1}{\Delta \omega}  \int_{\omega_a}^{\omega_b} \frac{{\rm d} \Re \chi_{\textrm{opt}}(\omega)}{{\rm d} \omega}  \,{\rm d}\omega &= \frac{\omega_p^2}{\Delta \omega} \int_{\omega_a}^{\omega_b} \frac{2\omega}{(\omega_0^2 - \omega^2)^2} \,{\rm d}\omega \nonumber \\
&= \frac{\omega_p^2}{\Delta \omega} \left( \frac{1}{ {\omega_0}^2 - \omega_b^2} - \frac{1}{ {\omega_0}^2 - \omega_a^2}  \right) \nonumber \\
&= \frac{\omega_p^2}{\Delta \omega} \frac{\left( \omega_b^2 - \omega_a^2 \right)}{\left( {\omega_0}^2 - \omega_a^2 \right) \left( {\omega_0}^2 - \omega_b^2 \right)} = \overline{\chi'}. \label{eq:avgdisp}
\end{align}
To simplify \eqreftwo{avgsus}{avgdisp} and obtain an analytical bound in terms of dispersion, we assume that the material is not too dispersive in the frequency interval of interest such that $\omega_a, \omega_b \ll \omega_0$. In this limit, defining the center frequency $\omega_c \equiv \frac{\omega_a + \omega_b}{2}$, the above expressions reduce to:
\begin{align}
\frac{1}{\Delta \omega}  \int_{\omega_a}^{\omega_b} \Re \chi_{\textrm{opt}}(\omega) \,{\rm d}\omega \approx \frac{\omega_p^2}{\omega_0^2}, \label{eq:smsus}
\end{align}
\begin{align}
\frac{1}{\Delta \omega}  \int_{\omega_a}^{\omega_b} \frac{{\rm d} \Re \chi_{\textrm{opt}}(\omega)}{{\rm d} \omega}  \,{\rm d}\omega \approx \frac{2\omega_p^2\omega_c}{\omega_0^4}= \overline{\chi'}. \label{eq:smdisp}
\end{align}
The equations above can be combined to bound the average susceptibility in terms of average cutoff dispersion $\overline{\chi'}$:
\begin{align}
\frac{1}{\Delta \omega}  \int_{\omega_a}^{\omega_b} \Re \chi(\omega) \,{\rm d}\omega \leq  \omega_p \sqrt{\frac{\overline{\chi'}}{2\omega_c}}. \label{eq:bndsus}
\end{align}
For nonzero bandwidth, using the average-susceptibility bound, \eqref{bndsus}, to bound the bandwidth-averaged refractive index is more subtle than the single-frequency case, since $ \int n^2(\omega) \,{\rm d}\omega \neq \left( \int n(\omega) \,{\rm d}\omega \right)^2$. To circumvent this issue, we use Cauchy--Schwartz inequality (along with the relation $n = \sqrt{1 + \chi}$) to obtain the following relation:
\begin{align}
\frac{1}{\Delta \omega}  \int_{\omega_a}^{\omega_b} n \,{\rm d}\omega \leq \left( \frac{1}{\Delta \omega} \int_{\omega_a}^{\omega_b} n^2 \,{\rm d}\omega \right)^{1/2} = \left( \frac{1}{\Delta \omega} \int_{\omega_a}^{\omega_b}\Re \chi \,{\rm d}\omega + 1 \right)^{1/2}. \label{eq:cauchy}
\end{align}
Combining \eqreftwo{bndsus}{cauchy} leads to the bound on bandwidth-averaged refractive index:
\begin{align}
\frac{1}{\Delta \omega}  \int_{\omega_a}^{\omega_b} n \,{\rm d}\omega \leq \left( \omega_p \sqrt{\frac{\overline{\chi'}}{2\omega_c}} + 1 \right)^{1/2} \label{eq:bndri}.
\end{align}
Unlike the single-frequency case, we cannot express the bound in terms of $\overline n'$, or the (maximum allowable) dispersion on $n$ averaged over bandwidth. This is because $\overline \chi'$ cannot be directly related to $\overline n'$ for nonzero bandwidth---the inequality $\overline \chi' \geq 2 \overline n'$ holds for $n \geq 1$, but this does not appear useful in transforming \eqref{bndri} in terms of $\overline n'$.

\section{Effect of $\Im \chi$ broadening on maximal refractive index}

Real materials are characterized by a smeared-out $\Im \chi$ spectrum, deviating far away from a sharp, localized resonance. In this section, we investigate to what extent does such smearing degrade our refractive-index bound. We restrict the analysis to a uniformly broadened $\Im \chi$ distribution for simplicity, i.e., a constant value within some prescribed region $[\omega_a, \omega_b]$ described by a bandwidth $\Delta \omega \equiv \omega_b - \omega_a$ with center frequency $\omega_0 \equiv \frac{\omega_a + \omega_b}{2}$, and zero elsewhere (the amplitudes are normalized in accordance with the sum rule, see \eqref{fsum}). We can derive analytical expressions for the susceptibility and dispersion as follows:
\begin{align}
\chi(\omega) &= \frac{\omega_p^2}{\Delta \omega} \int_{\omega_a}^{\omega_b} \frac{1}{\omega_i^2 - \omega^2} \,{\rm d}\omega_i \nonumber \\
&= \frac{\omega_p^2}{\Delta \omega} \frac{1}{\omega} \ln{ \left\{ \frac{ \left( \omega+\omega_a \right)\left( \omega_b - \omega \right) }{ \left( \omega+\omega_b \right)\left( \omega_a - \omega \right) } \right\} }, \label{eq:brchi}
\end{align}
\begin{align}
 \frac{{\rm d} \chi}{{\rm d} \omega}(\omega) &= \frac{2\omega_p^2 \omega}{\Delta \omega} \int_{\omega_a}^{\omega_b} \frac{1}{\left( \omega_i^2 - \omega^2 \right)^2} \,{\rm d}\omega_i \nonumber \\
&= \frac{\omega_p^2}{\Delta \omega} \frac{1}{\omega^2} \left[ \frac{\omega_b \omega}{\omega^2 - \omega_b^2}  -  \frac{\omega_a \omega}{\omega^2 - \omega_a^2} +\ln{ \left\{ \frac{ \left( \omega+\omega_a \right)\left( \omega_b - \omega \right) }{ \left( \omega+\omega_b \right)\left( \omega_a - \omega \right) } \right\} }  
 \right] \label{eq:brdisp}.
\end{align}

\begin{figure} [t!]
    \includegraphics[width=0.5\linewidth]{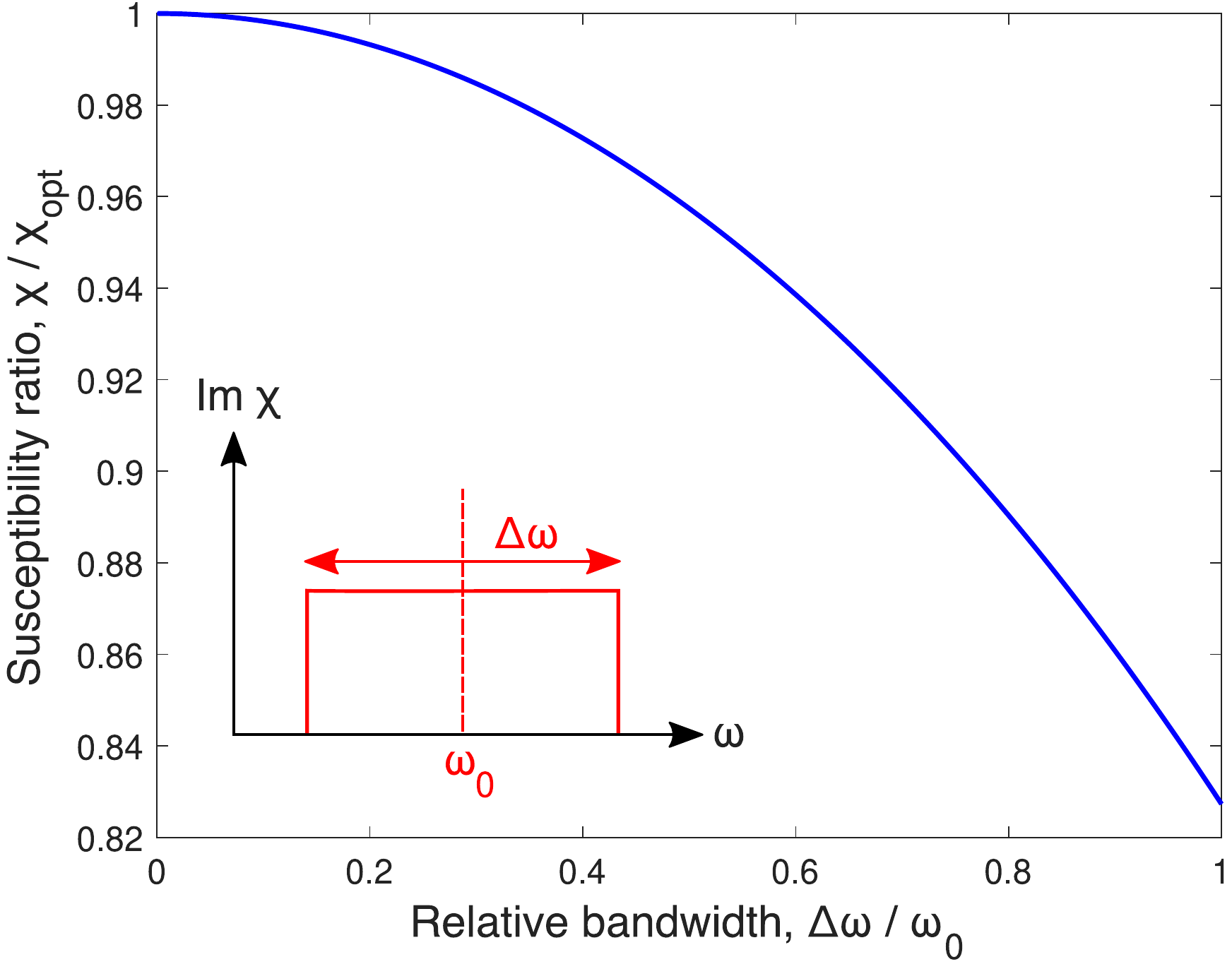} 
    \caption{Plot of susceptibility under different values of relative bandwidth, relative to its maximal value with the same dispersion (when the $\Im \chi$ distribution is a delta function). As the relative bandwidth increases, the susceptibility deviates further away from its maximum. The bottom left schematic shows the uniform $\Im \chi$ distribution considered here. The frequency of interest is set to $\omega = 0.1\omega_p$ and the center frequency of $\Im \chi$ distribution $\omega_0 = \omega_p$, but the result is unchanged as long as the $\Im \chi$ spectrum is far away from the frequency of interest, i.e., $\omega_0 - \Delta \omega /2 \gg \omega$.}  
    \label{fig:broad_imchi} 
\end{figure} 		

To quantity the extent to which $\Im \chi$ broadening reduces the maximal susceptibility, we compute the ratio of the susceptibility \eqref{brchi} to its optimal value (\eqref{sfchi_opt} with dispersion given by \eqref{brdisp}). As shown in \figref{broad_imchi}, the larger the broadening as measured by relative bandwidth $\Delta \omega / \omega_0$, the smaller the susceptibility relative to its maximal value with the same dispersion (since the refractive index $n = \sqrt{1+\chi}$, the ratio of $n$ to its maximum value with the same dispersion will be affected to a lesser extent by $\Im \chi$ broadening, especially more so when the susceptibility and dispersion are small). In the limit of vanishing relative bandwidth, the $\Im \chi$ distribution converges to a delta function and thus we obtain the optimal susceptibility as expected. 

Also, the ratio of susceptibility to its optimal value depends only on the relative bandwidth $\Delta \omega / \omega_0$, instead of both $\Delta \omega$ and $\omega_0$ separately, as long as the $\Im \chi$ spectrum is far enough from the frequency of interest $\omega$. To see this, rescale the $\Im \chi$ spectrum such that it lies in $[\alpha \omega_a, \alpha \omega_b]$ for some positive number $\alpha$. In that case, it can be shown that both the susceptibility in \eqref{brchi} and the optimal susceptibility in \eqref{sfchi_opt} (with dispersion given by \eqref{brdisp}) scale as $\sim 1/\alpha^2$ as long as $\omega_a, \omega_b \gg \omega$. Thus, their ratio is independent of $\alpha$, consistent with the fact that the relative bandwidth is unchanged by such scaling.

\section{Refractive-index bound based on alternative Kramers--Kronig relations}

Previously, our starting point for establishing the maximal refractive index was Kramers--Kronig relations for susceptibility, explicitly written out in \eqref{kk}. This was the defining relation from which we could obtain a general representation of susceptibility. However, there exists an analogous relation based on the refractive index $n$ instead of susceptibility $\chi$~\cite{King1976}: 
\begin{align}
\Re n(\omega) = 1 + \frac{2}{\pi}  \int_0^\infty \frac{\omega' \Im n(\omega')}{\omega'^2 - \omega^2} \,{\rm d}\omega'.  \label{eq:nkk} 
\end{align}
Similar to the sum rule on the imaginary part of susceptibility, a corresponding sum rule for the imaginary part of refractive index is given as follows~\cite{King1976}:
\begin{align}
\int_0^\infty \omega \Im n(\omega) \,{\rm d}\omega = \frac{\pi \omega_p^2}{4}.  \label{eq:nsum} 
\end{align}
Notice that, apart from a factor-of-2 difference, \eqreftwo{nkk}{nsum} are identical in form with \eqreftwo{kk}{fsum}. Thus, we can follow the same procedure as in \secref{genrep} to construct a general representation of refractive index, where $\sum_{i=1}^N c_i = 1$:
\begin{align}
\Re n(\omega) =1 + \frac{1}{2} \sum_{i=1}^N  \frac{c_i \omega_p^2}{\omega_i^2 - \omega^2}.  \label{eq:nrep} 
\end{align}		

Following the steps detailed in \secref{proofosc}, the optimal solution only contains one term among the summation in \eqref{nrep} (the proof carries over after we divide $f_i(\omega)$, $g_i(\omega)$ by 2 and replace $\chi$ by $n-1$ in \eqref{opt_sf}). The optimal solution for maximal refractive index, given the largest allowed value of dispersion $n' \equiv \frac{{\rm d} n}{{\rm d} \omega}$, is (where $\omega_0$ is determined by the dispersion constraint):
\begin{align}
\Re n_{\textrm{opt}}(\omega) = 1+  \frac{\omega_p^2}{2}\frac{1}{\omega_0^2 - \omega^2}.	\label{eq:sfn}
\end{align}
Its dispersion will equal $n'$, the maximum allowed value:
\begin{align}
\frac{{\rm d} \Re n_{\textrm{opt}}(\omega)}{{\rm d} \omega} = \frac{\omega_p^2 \omega}{(\omega_0^2 - \omega^2)^2} = n'.	\label{eq:sfndisp}
\end{align}
We can thus bound the refractive index after substituting \eqref{sfndisp} into \eqref{sfn} (again, we drop the $\Re$ symbol since $n_{\textrm{opt}}$ is manifestly real):
\begin{align}
n \leq 1 + \frac{\omega_p}{2} \sqrt{\frac{n'}{\omega}}.	\label{eq:sfn_opt}
\end{align}

\section{Comparison of n-KK and $\chi$-KK bounds}
In the main text, we claim that the bound derived from the KK relations for $\chi$, which we refer to as $\nc$, is always tighter than the the bound derived from the KK relations for $n$, which we refer to as $\nn$. We repeat here the two bounds:
\begin{align}
    \frac{\left(\nc^2-1\right)^2}{\nc} = \frac{\omega_p^2 n'}{\omega},
    \label{eq:chiKK}
\end{align}
and
\begin{align}
    \nn = 1 + \frac{\omega_p}{2} \sqrt{\frac{n'}{\omega}}.
    \label{eq:nKK}
\end{align}
To compare the two bounds, we can take the second, subtract 1 from both sides, and then square them, to find:
\begin{align}
    \left(\nn - 1\right)^2 = \frac{\omega_p^2}{4}\left(\frac{n'}{\omega}\right).
    \label{eq:nKK2}
\end{align}
This is starting to look more like the $\chi$-KK bound of \eqref{chiKK}. We can rewrite the $\chi$-KK bound of \eqref{chiKK} by noting that $(n^2-1)^2 = (n-1)^2(n+1)^2$, in which case we have
\begin{align}
    \frac{\left(\nc-1\right)^2\left(\nc+1\right)^2}{\nc} = \frac{\omega_p^2 n'}{\omega},
\end{align}
or,
\begin{align}
    \left(\nc - 1 \right)^2 = \frac{\omega_p^2 n'}{\omega} \left( \frac{\nc}{\left(\nc+1\right)^2} \right),
\end{align}
which now has the same left-hand side as the $n$-KK bound of \eqref{nKK2}. Note that both bounds of \eqref{chiKK} and \eqref{nKK} imply refractive index bounds greater than 1, for positive dispersion $n'$. Given that the bounds are greater than 1, we can write $\nc = x+1$, where $x$ is positive, and note that $\nc/(\nc+1)^2 < 1/4$, by substitution of $x$. Hence,
\begin{align}
    \left(\nc - 1 \right)^2 < \frac{\omega_p^2 n'}{4\omega} = \left(\nn - 1\right)^2.
\end{align}
Thus we have shown that
\begin{align}
    \nc < \nn,
\end{align}
and, indeed, the $\chi$-KK bound is always smaller than the $n$-KK bound.

\section{Maximal refractive index allowing small-but-nonzero losses}

In this section, we analyze to what extent the refractive-index bound established in the main text changes when we allow for small, but nonzero, losses at the frequency of interest (where we maximize refractive index). This involves two steps: 1) we find the maximal susceptibility allowing nonzero losses, and 2) relate the maximal susceptibility to maximal refractive index, which is nontrivial when we allow for lossy media at the frequency of interest.

One way to allow for losses at the frequency of interest is to modify the general representation of susceptibility (\eqref{sumosc}) by introducing a nonzero loss rate $\gamma$ into the oscillators: 
\begin{align}
\Re \chi(\omega) = \sum_{i=1}^N  \frac{c_i \omega_p^2}{\omega_i^2 - \omega^2 - i \gamma \omega}.  \label{eq:sumloss} 
\end{align}	
where now the oscillators are spaced $2\gamma$ from one another (i.e. the oscillator frequencies $\omega_i$ are at a distance $2\gamma$ from adjacent ones). By artificially adding a nonzero linewidth to each oscillators, \eqref{sumloss} enables us to quantify the effect of loss on maximal refractive index. 

\begin{figure} [t!]
    \includegraphics[width=0.5\linewidth]{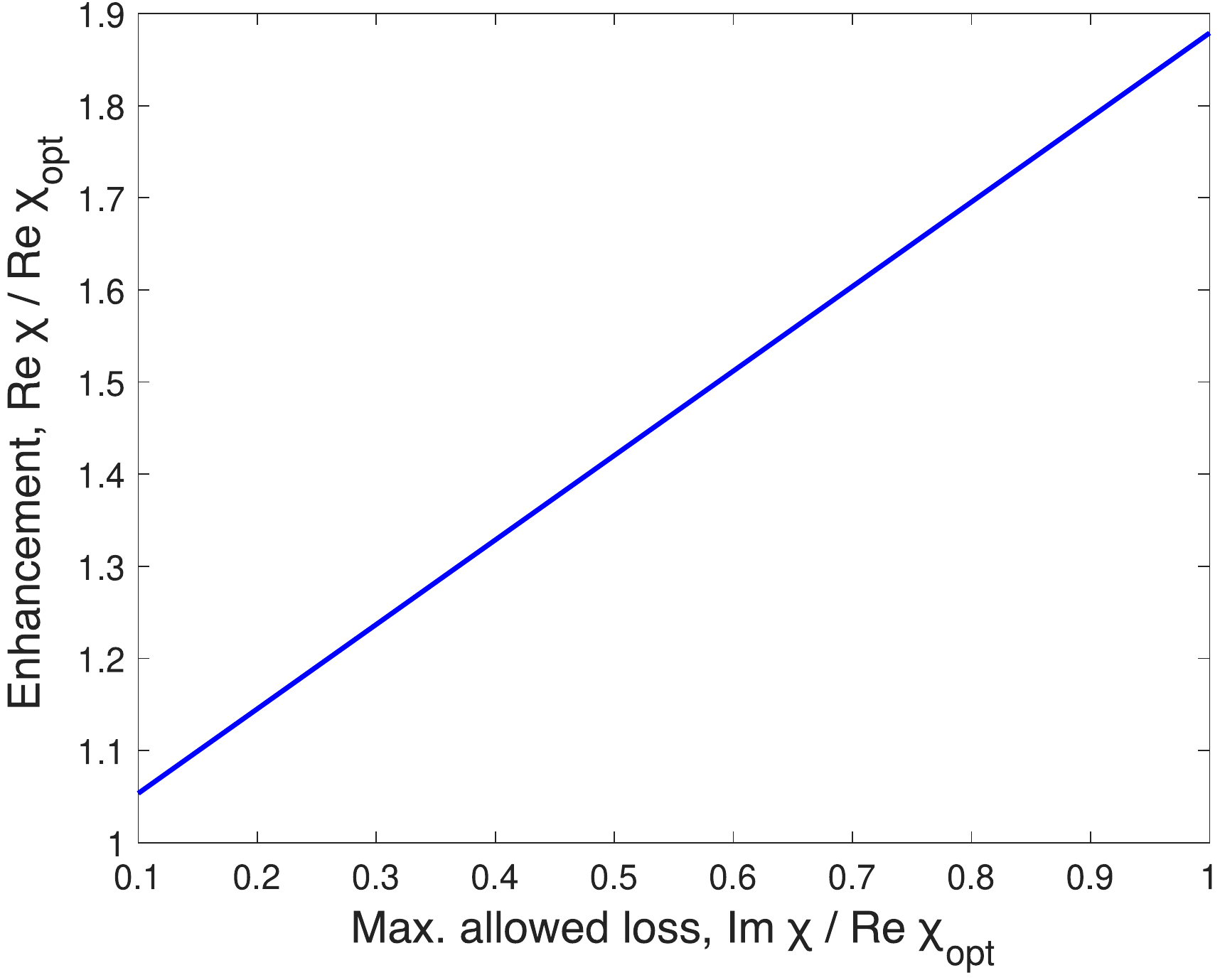} 
    \caption{Enhancement in maximal susceptibility, relative to that for transparent media ($\Re \chi_{\textrm{opt}}$), as a function of maximum allowable loss (measured relative to $\Re \chi_{\textrm{opt}}$). As we allow for larger losses, the maximal susceptibility can surpass that for transparent media, albeit by a modest amount. Here, we fix the frequency $\omega$ at $2$ eV and the loss rate of each oscillator $\gamma = 0.25$ eV, such that the oscillators are spaced at $0.5$ eV from one another (the oscillator nearest to the frequency of interest has $\omega_i = 2.13$ eV).}  
    \label{fig:loss_RI} 
\end{figure} 	

In \figref{loss_RI}, we optimize the real part of susceptibility at $2$ eV subject to maximum allowable dispersion of GaAs~\cite{Aspnes1986} (or equivalently, assuming a transparency window of $\approx 4.7$ eV centered around $2$ eV---what matters for our purposes is the maximum dispersion at $2$ eV). As shown by the figure, the maximal susceptibility increases linearly with the maximum allowed loss ($\Im \chi$) at the same frequency. Relative to the maximal susceptibility for transparent media, $\Re \chi_{\textrm{opt}}$ (\eqref{sfchi_opt}), the maximal susceptibility allowing for losses is enhanced only by a negligible amount, not surpassing a factor of two even when we allow for losses comparable to $\Re \chi_{\textrm{opt}}$ itself. Intuitively, as one allows for larger loss at the frequency of interest, the oscillator strengths $c_i$ are redistributed such that the net contribution to (the real part of) susceptibility increases in proportion (although not shown in \figref{loss_RI}, the maximal susceptibility saturates at $\Re \chi \approx 4.15 \Re \chi_{\textrm{opt}}$ at a loss level of $\Im \chi \approx 3.75\Re \chi_{\textrm{opt}}$). 

To translate the above results in terms of maximal refractive index, we use the following definitions linking refractive index to susceptibility (allowing for loss):
\begin{align}
n^2 - k^2 = \Re \chi + 1  \nonumber \\
2nk = \Im \chi.  \label{eq:nkdef} 
\end{align}
Combining \eqref{nkdef}, we obtain the following relation on refractive index:
\begin{align}
n^2 = \frac{1}{2} \left(1+ \Re \chi \pm \sqrt{1 + \Re \chi + \Im \chi } \right) .  \label{eq:n2loss} 
\end{align}
We restrict ourselves to small losses (i.e., $\Im \chi \ll \Re \chi$), which allows us to perform Taylor expansion with respect to the (small) parameter $\Im \chi / \Re \chi$. Thus, Taylor-expanding \eqref{n2loss} (taking the positive sign):
\begin{align}
n^2 = 1+ \Re \chi \left( 1 + \frac{1}{4}  \frac{(\Im \chi)^2}{( 1 + \Re \chi)^2} \right),  \label{eq:n2taylor} 
\end{align}
and taking its square root, we can Taylor-expand again to obtain the refractive index:
\begin{align}
n \approx \sqrt{1 +\Re \chi } \left( 1 + \frac{1}{8} \frac{(\Im \chi)^2}{( 1 + \Re \chi)^2} \right).  \label{eq:nsimp} 
\end{align}
Thus, in the limit of small loss, we can simply drop the second-order term above to obtain:
\begin{align}
\frac{n}{n_{\textrm{opt}}} \approx \sqrt{ \frac{1+\Re \chi}{1+\Re \chi_{\textrm{opt}}} }.  \label{eq:nratio} 
\end{align}
\Eqref{nratio} says that the refractive-index enhancement is even smaller than that for the susceptibility, which is already negligible for small losses. Thus, dramatic increases in maximal refractive index is not possible by allowing for small, but nonzero, losses.

\section{Derivation of bandwidth-based bound} \label{sec:bwbound}

Recall that the upper bound on the minimal refractive index averaged over some bandwidth $\Delta \omega$ centered about $\omega$ solves the following optimization problem: 
\begin{equation}
    \begin{aligned}
        & \underset{c_i}{\text{maximize}} & &  \text{min} \ \frac{1}{\Delta \omega}\int_{\omega - \frac{\Delta \omega}{2}}^{\omega + \frac{\Delta \omega}{2}} \Re \chi(\omega') \,{\rm d}\omega'  = \sum_{i=1}^N  c_i \, h_i(\omega)  \\
        & \text{subject to}       & & \Im \chi(\omega')= 0 \quad \textrm{for} \quad \omega' \in [\omega - \frac{\Delta \omega}{2}, \omega + \frac{\Delta \omega}{2}], \\
        & & & \sum_{i=1}^N c_i  = 1, \ c_i \geq 0.  \label {eq:opt_bw}
    \end{aligned}
\end{equation}
where $h_i(\omega) = \frac{1}{\Delta \omega} \int_{\omega - \frac{\Delta \omega}{2}}^{\omega + \frac{\Delta \omega}{2}} \frac{\omega_p^2}{\omega_i^2 - \omega'^2}\,{\rm d}\omega' =  \frac{\omega_p^2}{\omega_i \Delta \omega} \left( \tanh^{-1}\left(\frac{\omega + \frac{\Delta \omega}{2}}{\omega_i}\right) - \tanh^{-1}\left(\frac{\omega - \frac{\Delta \omega}{2}}{\omega_i}\right) \right)$ (the last expression is valid for $\omega_i > \omega+  \frac{\Delta \omega}{2}$).  

While \eqref{opt_bw} looks complicated, the optimal solution can be deduced from straightforward inspection and, analogous to the problem with dispersion constraint, is a single oscillator (i.e. $c_i =0$ for all $i$ except when $\omega_i = \omega + \frac{\Delta \omega}{2}$). To see this, notice that the constraint in \eqref{opt_bw} is equivalent to restricting the coefficients $c_i$ to vanish in the interval $[\omega - \frac{\Delta \omega}{2}, \omega + \frac{\Delta \omega}{2}]$, since $c_i$ is proportional to $\Im \chi$ at the same frequency---see \eqref{loc} (strictly speaking, the oscillator frequency is $\omega + \frac{\Delta \omega}{2} + \lim_{\delta \rightarrow 0^{+}} \delta $ to avoid loss at $\omega + \frac{\Delta \omega}{2}$, but this has no bearing on the following results). Thus, the optimal susceptibility profile is a single oscillator with oscillator frequency closest to the frequency band of interest from above, which is at $\omega_i = \omega + \frac{\Delta \omega}{2}$, and is given as follows:
\begin{align}
\Re \chi_{\textrm{opt}}(\omega') = \frac{\omega_p^2}{(\omega + \frac{\Delta \omega}{2})^2 - \omega'^2},	\label{eq:bwchi}
\end{align}
where we use the symbol $\omega'$ (which can assume any value) to distinguish it from the center frequency of interest $\omega$. 
Since $\Re \chi_{\textrm{opt}}$ is monotonically increasing with frequency, we know that the susceptibility, and hence refractive index, is smallest at $\omega' = \omega - \frac{\Delta \omega}{2}$. Thus, we can bound the minimal refractive index $n_{\textrm{min}}$ using the relation $n = \sqrt{1 + \Re \chi}$ (since $\chi$ is real away from $ \omega + \frac{\Delta \omega}{2}$): 
\begin{align}
n_{\textrm{min}} &\leq \sqrt{1 + \Re \chi_{\textrm{opt}}(\omega - \frac{\Delta \omega}{2})} \nonumber \\ 
&= \sqrt{1 + \frac{\omega_p^2}{2\omega \Delta\omega}}.	\label{eq:bwnopt}
\end{align}
We can further simplify \eqref{bwnopt} by recognizing that $\omega_p \gg \omega, \Delta\omega$ for practical cases of interest (for typical materials, $\omega_p$ hovers around $10\sim30$ eV, whereas $\omega$ is at most $\sim 5$ eV---one is often interested in visible or lower frequencies---and $\Delta \omega$ is even smaller than $\omega$):
\begin{align}
n_{\textrm{min}} \leq \frac{\omega_p}{\sqrt{2\omega \Delta\omega}}.	\label{eq:bwnopt_simple}
\end{align}

Apart from bounding $n_{\textrm{min}}$, we can also bound the refractive index at the center frequency $\omega$, which is in fact used to derive the bound on group index in \secref{gind}:
\begin{align}
n(\omega) &\leq \sqrt{1 + \Re \chi_{\textrm{opt}}(\omega) } \nonumber \\ 
&= \sqrt{1 + \frac{\omega_p^2}{\frac{\Delta \omega}{2} (\omega + \frac{\Delta \omega}{2})}} \nonumber \\
&\leq \frac{2 \omega_p}{\sqrt{\Delta\omega (4\omega + \Delta\omega)}},	 \label{eq:bwnopt_center}
\end{align}
where we have used \eqref{bwchi} in going from the first to second line, and used the fact that $\omega_p \gg \omega, \Delta\omega$ for practical purposes to go from the second to the last line (note that the group-index bound in \eqref{nbound_group} of \secref{gind} uses the symbol $\delta$ instead of $\Delta \omega$, as the latter there denotes the bandwidth over which the group index is averaged).

\section{Kramers--Kronig relations and $f$-sum rule for nonlocal permittivity tensor}

For completeness, we explicitly write the Kramers--Kronig relations and the $f$-sum rule for the general case of permittivity tensor $\epsnl$, allowing for nonlocality. From Ref.~\cite{Altarelli1972} (Eq. (28) and (30) therein), the most general form of Kramers--Kronig relations and the $f$-sum rule are as follows (expressed in terms of $\chi_{ij}(\omega,\mathbf{k}) \equiv \varepsilon_{ij}(\omega,\mathbf{k}) - \delta_{ij}$):
\begin{align}
\Re \chi_{ij}(\omega,\mathbf{k}) = -\frac{1}{\pi \omega}  \int_0^\infty \left( \Im \chi_{ij}(\omega',\mathbf{k}) - \Im \chi_{ij}(\omega',-\mathbf{k}) \right) \,{\rm d}\omega' -\frac{1}{\pi \omega}  \int_0^\infty \left( \frac{\Im \chi_{ij}(\omega',\mathbf{k})}{\omega - \omega'} + \frac{\Im \chi_{ij}(\omega',-\mathbf{k})}{\omega + \omega'} \right) \omega' \,{\rm d}\omega',   \label{eq:kkgen} 
\end{align}

\begin{align}
\int_0^\infty \omega  \left( \Im \chi_{ij}(\omega,\mathbf{k}) + \Im \chi_{ij}(\omega,-\mathbf{k}) \right) \,{\rm d}\omega = \pi \omega_p^2 \delta_{ij}.   \label{eq:sumrulegen} 
\end{align}

For reciprocal media, we can use the relation $\chinl^T = \overline{\overline{\chi}}(\omega,-\mathbf{k})$~\cite{Agranovich2013} to simplify \eqreftwo{kkgen}{sumrulegen} when $i = j$ (diagonal components). The Kramers--Kronig relations, exactly analogous to \eqref{kk} for local, scalar permittivity, then hold for each diagonal component of $\chinl$ and each wavevector $\mathbf{k}$ (where the integral is a principal-valued integral):
\begin{align}
\Re \chi_{ii}(\omega,\mathbf{k}) = \frac{2}{\pi}  \int_0^\infty \frac{\omega' \Im \chi_{ii}(\omega',\mathbf{k})}{\omega'^2 - \omega^2} \,{\rm d}\omega'.   \label{eq:kknl} 
\end{align}

Similarly, the $f$-sum rule holds for each component of $\epsnl$ (analogous to \eqref{fsum}):
\begin{align}
\int_0^\infty \omega  \Im  \chi_{ii}(\omega,\mathbf{k}) \,{\rm d}\omega = \frac{\pi}{2}\omega_p^2 \delta_{ii}.   \label{eq:sumrulenl} 
\end{align}

It should be emphasized again that the indices in $\chi_{ii}$ refer to arbitrary diagonal component and in no way implies summation. These relations imply that our previous analysis based on local, isotropic susceptibility carries over to each diagonal component of nonlocal, reciprocal, anisotropic tensor $\epsnl$, and will be useful later when we bound the refractive index of generic bianisotropic media.
%
%

\section{Maximal refractive index for permittivity tensor}

In this section, we derive the refractive index for a general (possibly nonlocal) permittivity tensor $\epsten$ (for brevity, implicitly assumed to be evaluated at a fixed frequency $\omega$ and wavevector $\mathbf{k}$ throughout this section). We start from Maxwell's equation in frequency space (we assume magnetic permeability to be unity, since any bianisotropic media can be described by an effective permittivity tensor):
\begin{align}
\nabla \times \mathbf{H} = i \omega \epsten \mathbf{E} \nonumber \\
\nabla \times \mathbf{E} = -i \omega \mu_0 \mathbf{H}.
\end{align}
For plane waves of the form $\mathbf{E} = \mathbf{e_0} e^{ik \cdot r}, \mathbf{H} = \mathbf{h_0} e^{ik \cdot r}$, the above equations reduce to follows:
\begin{align}
\mathbf{k} \times \mathbf{h_0} = \omega \epsten \mathbf{e_0} \nonumber \\
\mathbf{k} \times \mathbf{e_0} = - \omega \mu_0 \mathbf{h_0}.
\end{align}
Combining the two equations above leads to the following:
\begin{align}
\mathbf{k} \times \mathbf{k} \times \mathbf{e_0} = - \omega^2 \mu_0 \epsten \mathbf{e_0}. \label{eq:twocurl}
\end{align}
Let us define the wavevector as $\mathbf{k} = \frac{n\omega}{c}\mathbf{\hat{s}}$, where the refractive index $n$ is not necessarily defined along different axes (for e.g., the polarization eigenstates are circularly polarized waves for gyrotropic or chiral media). Using the identity $\mathbf{k} \times \mathbf{k} \times \mathbf{e_0} = \mathbf{k} (\mathbf{k} \cdot \mathbf{e_o}) - |\mathbf{k}|^2 \mathbf{e_0}$, \eqref{twocurl} can be simplified as follows:
\begin{align}
n^2 \left[ \mathbf{\hat{s}} ( \mathbf{\hat{s}} \cdot \mathbf{e_0}) - \mathbf{e_0} ) \right]= - \epsten \mathbf{e_0}. \label{eq:eiglike}
\end{align}
We can rewrite \eqref{eiglike} as a generalized eigenvalue problem, where $B = \id - \mathbf{\hat{s}}\mathbf{\hat{s}}^T$ and $\mathbf{e_0}$ denotes the eigenvector corresponding to the electric-field polarization (in the main text, $\epsten$ refers to the nonlocal effective permittivity tensor $\epseff$):
\begin{align}
\epsten \mathbf{e_0} = n^2 B \mathbf{e_0}. \label{eq:geneig} 
\end{align}
We can choose to work in a polarization basis in which $\epsten$ is diagonal at the frequency of interest. Since we are interested in lossless materials, we can take $\epsten$ to be Hermitian; combined with the diagonality, this implies that $\epsten$ is real-symmetric, as is $B$ for propagating waves. Hence we can consider real-valued quantities for $\mathbf{e_0}$ and $n^2$.

There is a trivial solution to \eqref{geneig}, which is that the refractive index would appear to be able to go to infinity. This is because $B$ is a rank-two matrix (it projects onto the subspace of the two directions orthogonal to $\mathbf{\hat{s}}$), and if one simply chooses $\mathbf{e_0}$ to be parallel to $\mathbf{\hat{s}}$, then \eqref{geneig} is satisfied for. However, we must also pay attention to the transversality constraint, $\nabla \cdot \mathbf{D} = 0$, which for a propagating wave can be written 
\begin{align}
    \mathbf{\hat{s}} \cdot \epsten \mathbf{e_0} = 0.
    \label{eq:tc}
\end{align}
If $\epsten$ is definite (either positive definite or negative definite), it is not possible for the field polarization vector $\mathbf{e_0}$ to equal the eigenvector with infinite eigenvalue, $\mathbf{\hat{s}}$, because that would imply that $\mathbf{\hat{s}}^T \epsten \sv = 0$, which is not possible for a positive definite real-symmetric matrix. Hence that solution is disqualified for definite $\epsten$. (We return to the indefinite case below.)

For definite $\epsten$, then, the largest possible refractive index is the solution of the Rayleigh quotient of \eqref{geneig} over all vectors not parallel to $\mathbf{\hat{s}}$:
\begin{align}
    (n^2)_{\textrm{max}} = \max_{u \neq \sv} \frac{u^T \epsten u}{u^T B u}. \label{eq:maxeig} 
\end{align}
Thanks to the non-parallel requirement, we can write $u = a\sv + \vv$, where $\vv$ is a vector with nonzero amplitude and which is orthogonal to $\sv$. If we insert this solution into the Rayleigh quotient, we have:
\begin{align}
    (n^2)_{\textrm{max}} = \max_{a,v^T \sv = 0} \frac{\vv^T \epsten \vv + 2 a \sv^T \epsten \vv + a^2 \sv^T \epsten \sv}{\vv^T B \vv}. \label{eq:maxeig2} 
\end{align}
Note from the transversality constraint, \eqref{tc}, if $\mathbf{e_0} = \vv + a\sv$, then it must be true that
\begin{align}
    a \sv^T \epsten \sv = - \sv^T \epsten \vv,
\end{align}
and so the maximum refractive index is given by
\begin{align}
    (n^2)_{\textrm{max}} = \max_{a,v^T \sv = 0} \frac{\vv^T \epsten \vv - a^2 \sv^T \epsten \sv}{\vv^T B \vv}. \label{eq:maxeig3} 
\end{align}
The denominator, $v^T B v$, must equal 1, as $B$ is a projection matrix and $\vv$ lies entirely in the subspace of the projection. If the permittivity tensor $\epsten$ is positive definite, then the second term in the numerator is negative and can only reduce the maximum index. If $\epsten$ is negative definite, then $n^2$ would be negative, we would actually be interested in maximizing the absolute value of the numerator of \eqref{maxeig3}, and again the second term could only result in a reduction. Thus, if $\epsten$ is definite, then we have that the largest possible magnitude of $n^2$ is given by the largest magnitude of the eigenvalue of $\epsten$. Since we have chosen a basis in which $\epsten$ is diagonal at the frequency of interest, with the diagonal terms denoted by
\begin{align}
\epseff = \begin{pmatrix}
                        \epsix & & \\
                        & \epsiy & \\
                        & & \epsiz  
                    \end{pmatrix},  \label{eq:diageffeps}
\end{align}
where the diagonal components are all real, then the eigenvalues are the diagonal components themselves. The maximum positive square of the refractive index is then given by:
\begin{align}
(n^2)_{\textrm{max}} \leq \max \{ \epsix,\epsiy,\epsiz \}, \label{eq:maxeigsimp} 
\end{align}
where $\epsix,\epsiy,\epsiz$ refer to the diagonal components in \eqref{diageffeps} and the equality holds only if $\max \{ \epsix,\epsiy,\epsiz \}$ belongs to the subspace projected onto by $B$.  

Since each diagonal component of arbitrary nonlocal permittivity tensor (recall $\epsten$ is evaluated at a fixed frequency $\omega$ and wavevector $\mathbf{k}$) obeys both the Kramers--Kronig relations and the $f$-sum rule, the largest eigenvalue of $\epsten$ (and hence the maximal refractive index for $\epsten$) can be written as a sum of lossless Drude--Lorentz oscillators as in \eqref{sumosc}. Thus, the refractive index for $\epsten$ is subject to the same refractive-index bound established earlier (see \eqref{nbound}), but with dispersion corresponding to the maximum principal component of $\epsten$. 

However, as mentioned above, the refractive index described by $\epsten$ can be arbitrarily large when $\epsten$ is indefinite, i.e. for the class of hyperbolic metamaterial permittivity tensors. This can be verified numerically, but we can provide physical intuition behind such behavior. For concreteness, let us take $\mathbf{\hat{s}}= \mathbf{\hat{z}}$ (i.e., pointing along the $z$-axis), for which $B = \textrm{diag}(1, 1, 0)$. Writing the eigenvector as $\mathbf{e_0} = ( e_x, e_y, e_z)^T$, 
\begin{align}
\epsten \begin{pmatrix}
        e_x \\
        e_y \\
        e_z
    \end{pmatrix} = n^2 \begin{pmatrix}
        e_x \\
        e_y \\
        0
    \end{pmatrix}. \label{eq:eigz} 
\end{align}
Straightforward inspection of \eqref{eigz} shows that $|e_z| \gg |e_x|, |e_y|$ is necessary in order to have gigantic refractive index ($n \gg \varepsilon_{ij}$ for all $i,j=x,y,z$). We can hence explicitly write down \eqref{eigz} as follows: 
\begin{align}
\begin{pmatrix}
        \varepsilon_{xx} e_x + \varepsilon_{xy} e_y + \varepsilon_{xz} e_z \\
        \varepsilon_{yx} e_x + \varepsilon_{yy} e_y + \varepsilon_{yz} e_z \\
         \varepsilon_{zx} e_x + \varepsilon_{zy} e_y + \varepsilon_{zz} e_z 
    \end{pmatrix} \approx \begin{pmatrix}
        \varepsilon_{xz} e_z \\
       \varepsilon_{yz} e_z \\
         \varepsilon_{zz} e_z 
    \end{pmatrix} = n^2 \begin{pmatrix}
        e_x \\
        e_y \\
        0
    \end{pmatrix}. \label{eq:eigfull} 
\end{align}
Thus, we obtain two conditions for gigantic refractive index: (1) $|e_x|$ is smaller than $|e_z|$ by a factor of about $\frac{ | \varepsilon_{xz} | }{n^2} \ll 1$, and similarly for $|e_y|$, (2) $|\varepsilon_{zz} e_z | \leq |\varepsilon_{zz}| \rightarrow 0$ (at least relative to other $\varepsilon_{ij}$). Now, we wish to show that $\epsten$ is indefinite for eigenvector $\mathbf{e_0}$ satisfying such conditions. To simplify the analysis, we can drop $\varepsilon_{xx}, \varepsilon_{xy}, \varepsilon_{yy}$ without affecting \eqref{eigfull}. In that case, we can directly compute the eigenvalues of $\epsten$ (denoted by $\lambda$):
\begin{align}
\det \begin{pmatrix}
        -\lambda & 0 & \varepsilon_{xz}  \\
        0 & -\lambda & \varepsilon_{yz} \\
        \varepsilon_{xz} & \varepsilon_{yz} & \varepsilon_{zz} - \lambda
    \end{pmatrix} = 0 \nonumber \\
    \lambda \left[ \lambda^2 - \varepsilon_{zz} \lambda - (\varepsilon_{xz}^2 + \varepsilon_{yz}^2) \right]=0. \label{eq:eigvalz} 
\end{align}
Since $\varepsilon_{xz}^2 + \varepsilon_{yz}^2 > 0$, \eqref{eigvalz} shows that the two eigenvalues are of the opposite sign (the remaining being 0). This demonstrates that $\epsten$ has to be indefinite for refractive index much greater than the (square root of) maximal eigenvalue of $\epsten$ (for a positive- or negative-semidefinite $\epsten$, finding the refractive index reduces to an eigenvalue problem involving only the $x,y$-components in \eqref{eigz}). While we have assumed the upper $2\times2$ submatrix of $\epsten$ to be zero for simplicity, the same conclusion holds even if we allow for arbitrary $\epsten$, but still require gigantic refractive index.  We have also taken the wavevector to be along the $z$-axis here, but our results here are also valid for other polarization eigenstates. 

\section{Conditions for positive-definite effective permittivity tensor}

In this section, we explicitly write down the nonlocal effective permittivity $\epseff$ alluded to in the main text, and derive under what conditions it is positive definite. As shown in the previous section, the refractive index can in principle be arbitrarily large for indefinite media, where the eigenvalues of $\epseff$ for a given $\omega$ and $\mathbf{k}$ can be of opposite sign. Thus, our bound on a material's refractive index holds only if it can be described by a positive-definite $\epseff$ (strictly speaking, our bound is valid also for negative-definite $\epseff$, but they are unimportant for our purposes of obtaining large refractive index).

For a generic linear, bianisotropic medium described by the following constitutive relations,
\begin{align}
\begin{pmatrix}
        \Dv \\
        \Bv
    \end{pmatrix} = 
  \begin{pmatrix}
        \epsten  & \frac{1}{c} \xiten \\
        \frac{1}{c} \zetaten & \muten
    \end{pmatrix}  
    \begin{pmatrix}
        \Ev \\
        \Hv
    \end{pmatrix},
    \label{eq:biani} 
\end{align}
one can adopt alternative phenomenological constitutive relations in which the effects of microscopic currents are all embedded in the electric displacement $\Dv$. Thus, one can define an effective permittivity $\epseff$ that encodes all the constitutive parameters in \eqref{biani}~\cite{Silveirinha2007}:
\begin{align}
\epseff = \epsten - \xiten \cdot \muten^{-1} \cdot \zetaten + \left( \xiten \cdot \muten^{-1} \times \frac{\mathbf{k}}{\beta} - \frac{\mathbf{k}}{\beta} \times \muten^{-1} \cdot \zetaten \right) + \frac{\mathbf{k}}{\beta} \times \left( \muten^{-1} - \overline{\overline{\mathbf{I}}} \right) \times \frac{\mathbf{k}}{\beta}
    \label{eq:epseff} 
\end{align}
where $\beta = \frac{\omega}{c}$ is the wavenumber in the host medium (here taken to be vacuum). 

Even for isotropic media, the effective permittivity $\epseff$ is a tensor due to spatial dispersion associated with magnetic response. Thus, we need to be careful under what conditions is $\epseff$ positive definite (in which case, our refractive-index bound applies). We consider isotropic and anisotropic permittivity and permeability in what follows, deriving sufficient conditions for their effective permittivities to be positive definite. Lastly, we comment on generic bianisotropic media, where the effective permittivity need not be positive definite for all spatial wavevector $\mathbf{k}$.


\subsection{Isotropic permittivity and permeability}
In this case, $\epsten = \varepsilon \id$ and $\muten = \mu \id$ and \eqref{epseff} simplifies ($\varepsilon$ and $\mu$ are real, since we are interested in lossless materials):
\begin{align}
\epseff = \varepsilon \id + \frac{1}{\beta^2} \left( \frac{1}{\mu} -1 \right) \mathbf{k} \times \id \times \mathbf{k}
    \label{eq:iso1} 
\end{align}
To evaluate the second term in \eqref{iso1}, it is handy to use dyadic notation to express it as follows:
\begin{align}
\mathbf{k} \times \id \times \mathbf{k} & = \left( k_x \xu + k_y \yu + k_z \zu \right) \times \left( \xu\xu + \yu\yu + \zu\zu \right) \times \left( k_x \xu + k_y \yu + k_z \zu \right) \nonumber \\
& = \left( k_x \xu + k_y \yu + k_z \zu \right) \times \left( k_x \left(\zu\yu - \yu\zu \right) - k_y \left(\zu\xu - \xu\zu \right)  + k_z \left(\yu\xu - \xu\yu \right) \right) \nonumber \\
& = -(k_y^2 + k_z^2)\xu\xu - (k_x^2 + k_z^2)\yu\yu - (k_x^2 + k_y^2)\zu\zu + k_xk_y \left( \xu\yu + \yu\xu \right)... \nonumber \\
& \quad \, + k_xk_z \left( \xu\zu + \zu\xu \right) + k_yk_z \left( \yu\zu + \zu\yu \right).
    \label{eq:iso2} 
\end{align}
With the help of \eqref{iso2}, we can explicitly write \eqref{iso1} in full matrix notation:
\begin{align}
\epseff = \begin{pmatrix}
                        \varepsilon & & \\
                        & \varepsilon & \\
                        & & \varepsilon 
                    \end{pmatrix} + 
                    \frac{\mu -1}{\mu} \begin{pmatrix}
                        k_y^2 + k_z^2 & -k_xk_y & -k_xk_z \\
                        -k_xk_y & k_x^2 + k_z^2 & -k_yk_z \\
                        -k_xk_z & -k_yk_z & k_x^2 + k_y^2
                    \end{pmatrix}
    \label{eq:iso3} 
\end{align}
Now, the first term is (trivially) positive definite for any positive $\varepsilon$. The second term is positive semidefinite if $\frac{\mu -1}{\mu} \geq 0$. One way to show this is to use Sylvester's criterion, according to which a matrix is positive semidefinite if all of its principal minors are nonnegative. For example,
\begin{align}
\left( \frac{\mu -1}{\mu} \right)^2 \begin{vmatrix}
                      k_y^2 + k_z^2 & -k_xk_y \\
                        -k_xk_y & k_x^2 + k_z^2 
                    \end{vmatrix} = \left( \frac{\mu -1}{\mu} \right)^2 k_z^2  \left( k_x^2 + k_y^2 + k_z^2 \right) \geq 0, 
    \label{eq:iso4} 
\end{align}
and similarly for the $3\times 3$ principal minor (which turns out to vanish exactly):
\begin{align}
\left( \frac{\mu -1}{\mu} \right)^3 \begin{vmatrix}
                        k_y^2 + k_z^2 & -k_xk_y & -k_xk_z \\
                        -k_xk_y & k_x^2 + k_z^2 & -k_yk_z \\
                        -k_xk_z & -k_yk_z & k_x^2 + k_y^2
                    \end{vmatrix} = 0.
    \label{eq:iso5} 
\end{align}
Thus, the term containing $\mu$ is positive semidefinite for $\frac{\mu -1}{\mu} \geq 0$. Since the sum of positive definite and positive semidefinite matrix is itself positive definite, we have shown that $\epseff$ is positive definite for isotropic permittivity and permeability if $\varepsilon > 0 $ and $\mu \geq 1$ (or $\mu <0$), so this excludes certain diamagnetic materials.

To be more precise, energy considerations~\cite{lindell1995conditions} show that $\epsten$ and $\muten$ are positive definite for lossless media. So, the only requirement is is that $\mu \geq 1$, as long as we adhere to lossless materials. However, this is a \emph{sufficient}, but not \emph{necessary}, condition---$\epseff$ will remain positive definite even if $\mu < 1$ unless $\mathbf{k}$ becomes comparable to $\beta = \frac{\omega}{c}$ in magnitude (or $\varepsilon \rightarrow 0$).

\subsection{Anisotropic permittivity and permeability}
We now consider general, anisotropic tensors $\epsten$ and $\muten$. Since they are real-symmetric for lossless media~\cite{Landau2013}, we can work in a basis in which $\muten$ is diagonal for simplicity, with principal components $\mu_{x,y,z}$. Defining $a_{x} = \frac{\mu_x -1}{\mu_x} $ and likewise for $y,z$, we obtain from \eqref{epseff}:
\begin{align}
\epseff = \epsten + 
                    \begin{pmatrix}
                        a_y k_z^2 + a_z k_y^2 & -a_z k_xk_y & -a_y k_xk_z \\
                        -a_z k_xk_y & a_z k_x^2 + a_x k_z^2 & -a_x k_yk_z \\
                        -a_y k_xk_z & -a_x k_yk_z & a_x k_y^2 + a_y k_x^2
                    \end{pmatrix}.
    \label{eq:ani1} 
\end{align}
Again, mirroring the reasoning in the previous section, the second term can be shown to be positive semidefinite if $a_x,a_y,a_z \geq 0$. For positive-definite $\epsten$ and $\muten$ (true for lossless media~\cite{lindell1995conditions}), $\epseff$ is positive definite as long as $\mu_x, \mu_y, \mu_z \geq 1$. Even if these conditions are not strictly met, $\epseff$ is indefinite only under particular circumstances. One would need diamagnetic response ($\mu_i < 1$ for one or more $i \in \{x,y,z\}$) and high spatial dispersion ($|\mathbf{k}|$ comparable to or larger than $\beta$) in certain directions to break the positive-definiteness of $\epseff$.

\subsection{General bianisotropic media}

For generic bianisotropic media (allowing for gyrotropy/chirality etc.), there is no simple criteria that ensures $\epseff$ is positive definite unless we know the value of $\mathbf{k}$. To show this, let us consider a lossless magneto-optical material for simplicity, characterized by isotropic $\xiten = \left( \chi + i \kappa \right) \id$, $\zetaten = \xiten^{\dagger} = \left( \chi - i \kappa \right) \id$, and $\muten = \mu \id$. In this case, the linear term in $\mathbf{k}$ of \eqref{epseff} reduces to:
\begin{align} 
                  \frac{2\kappa}{\mu} \begin{pmatrix}
                        0 & -ik_z & ik_y \\
                        ik_z & 0 & -ik_x \\
                        -ik_y & ik_x & 0
                    \end{pmatrix}
    \label{eq:gyro1} 
\end{align}
which is indefinite, and so $\epseff$ need not be positive definite for all values of $\mathbf{k}$. Thus, it is difficult to identify sufficient conditions for $\epseff$ to be positive definite---unless there is no spatial dispersion ($\mathbf{k}=0$), in which case $\overline{\overline{\varepsilon_{\textrm{eff}}}}(\omega,\mathbf{k}=0)=  \epsten - \xiten \cdot \muten^{-1} \cdot \zetaten$ is positive definite for lossless media~\cite{lindell1995conditions}. That being said, it is reasonable to expect that $\epseff$ will be positive-definite for a wide range of constitutive and dispersion parameters, as long as the (possibly indefinite) terms like  \eqref{gyro1} are swamped by other positive-definite terms in the overall expression for $\epseff$.

\section{Bounds on bandwidth-averaged group index} \label{sec:gind}
Closely related to refractive index is the group index $n_g$, defined as the ratio of the speed of light to group velocity in a medium. Since group index is related to phase index through a frequency derivative, $n_g = d(n \omega) / {\rm d}\omega$, its average over a bandwidth $\Delta \omega \equiv \omega - \omega_0$ can be written:
\begin{align}
\overline{n}_g \equiv \frac{1}{\Delta \omega}  \int_{\omega_0}^{\omega} n_g \,{\rm d}\omega = \frac{n(\omega) \omega - n(\omega_0) \omega_0 }{\Delta \omega}.  \label{eq:groupn}
\end{align}
After integrating over a nonzero bandwidth, the resulting expression only contains refractive index without the dispersion term. This enables us to carry over our refractive-index bounds established in \secref{bwbound} to bound the bandwidth-averaged group index $\overline{n}_g$ (we use the bandwidth-based bound here, since the bandwidth constraint is more natural in this context). Defining $\delta$ as the bandwidth centered around $\omega$ where no losses are allowed (not to be confused with $\Delta \omega$, the bandwidth over which we average the group index in \eqref{groupn}), we obtain the following upper bound (see \eqref{bwnopt_center}): 
\begin{align}
\overline{n}_g \leq n(\omega) \frac{\omega}{\Delta \omega} \leq 2 \omega_p \frac{\omega}{\Delta \omega} \sqrt{\frac{1}{\delta (4\omega + \delta)}}.  \label{eq:nbound_group}
\end{align}
At first glance, \eqref{nbound_group} may seem to imply that group index can become arbitrarily large with frequency. However, this bound becomes loose as $\omega$ increases indefinitely while keeping the bandwidth $\Delta \omega$ fixed (or more precisely, when $n(\omega_0) \omega_0$ becomes comparable in magnitude to $n(\omega)\omega$). Thus, it is more accurate to keep the fractional bandwidth ${\Delta \omega}/{\omega}$ constant, in which case the maximal group index mainly depends on the square-root term in \eqref{nbound_group}. The group index is fundamentally limited by the frequency of interest $\omega$ as well as $\delta$, which quantifies how close $\omega$ is to a nearby resonance (where losses dominate).

\section{Identifying the ideal high-index material} \label{sec:lightheavy}
\begin{figure*} [t!]
    \includegraphics[width=0.9\linewidth]{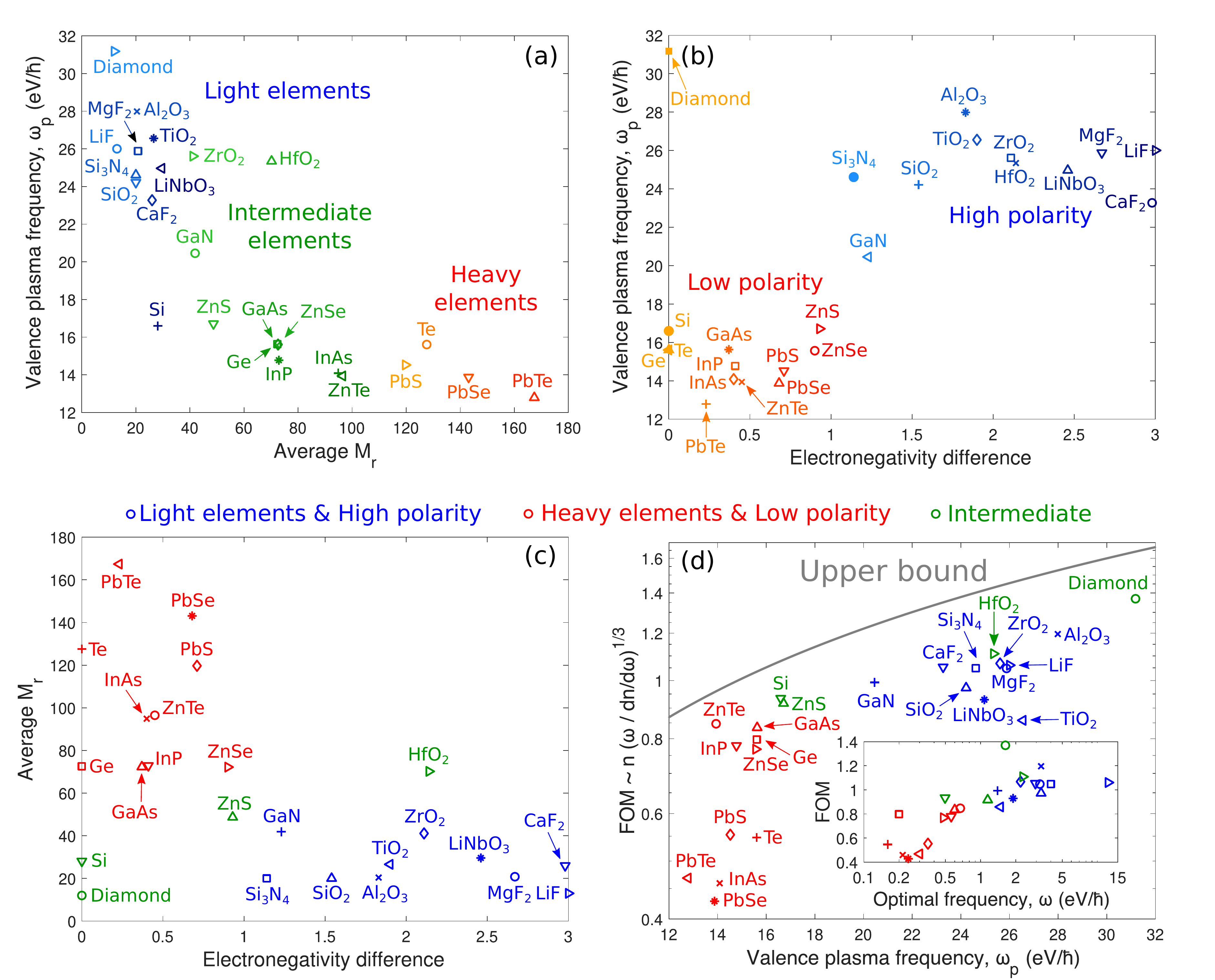} 
    \caption{Plasma frequency, based on valence electron density, plotted against (a) Average molar mass ($M_{\textrm{r}}$), defined as molar mass divided by the number of constituent atoms per molecule, and (b) Electronegativity difference ($\Delta$EN) for representative high-index materials. (c) Materials classified into three categories based on their average $M_{\textrm{r}}$ and $\Delta$EN---light elements $\&$ high polarity (blue), heavy elements $\&$ low polarity (red), and intermediate (green). (d) High-index materials evaluated by FOM $= \left[ \frac{(n^2 -1)^2 \omega}{n n'} \right]^{1/3} \approx  n \left(\frac{\omega}{n'} \right)^{1/3}$ for $n\gg1$ ($\frac{\omega}{n'}$ normalized to that of SiO$_2$ at 400 nm), compared to the bound. The inset shows the optimal frequencies at which the FOM is maximized for each material. Generally, materials with large plasma frequency are transparent in the UV or visible, while those with low plasma frequency are IR-transparent.}  
    \label{fig:lightvsheavy} 
\end{figure*} 
     
As demonstrated in the main text, many naturally occurring materials approach the bound at optical frequencies. Given that efforts to further approach the bound may not be very fruitful, another approach to improve the refractive index is to seek out materials with high plasma frequency $\omega_p$ (or equivalently, electron density $N_e$ since $\omega_p^2 = \frac{N_e e^2}{\varepsilon_0 m_e}$) as mentioned previously. This is already implicit in the expression for the upper bound in \eqref{nbound}, which shows that the maximal refractive index increases indefinitely with $\omega_p$ or $N_e$. More precisely, it scales with $\omega_p^{2/3}$ or $N_e^{1/3}$ for refractive indices much larger than unity (see \eqref{nbound_high}), which means that a twofold increase in electron density will increase maximal refractive index by a factor of $2^{1/3} \approx 1.26$. 

All the natural high-index materials across optical frequencies that we have seen in previous sections can be described by crystal lattice structures. For such structures, the valence electron density can be determined based on the number of atoms and their corresponding valence electrons within a unit cell (as mentioned in the main text, we use valence, instead of total, electron density as it gives a more accurate refractive-index bound over optical frequencies). Denoting the number of atoms per unit cell and average number of valence electrons per atom as $N_{\textrm{at}}$ and $N_{\textrm{val}}$ respectively (and the volume of unit cell as $V$), the valence electron density $N_e$ can be written as~\cite{Kaxiras2019}:
\begin{align}
N_e = \frac{N_{\textrm{at}} N_{\textrm{val}}}{V}. \label{eq:electron_density}
\end{align}
\Eqref{electron_density} demonstrates that, in order to maximize $N_e$, it is desirable to have as many atoms with large number of valence electrons within a small lattice volume. Based on these microscopic parameters, we identify two key physical quantities responsible for high electron density---average molar mass and electronegativity difference (explained further below)---that are easily measurable and abundant in literature.
		
Generally, molecules with high molar mass $M_{\textrm{r}}$ tend to occupy larger volume, while the average number of valence electrons per atom $N_{\textrm{val}}$ is mostly unaffected. Also, the greater the number of atoms in a molecule (a rough but convenient measure of $N_{\textrm{at}}$), the larger the electron density. Thus, materials with a small average $M_{\textrm{r}}$---molar mass divided by the number of constituent atoms per molecule---are conducive to high electron density, as demonstrated in \figref{lightvsheavy}(a). Another factor that sets the volume in crystals is the bond strength between molecules. Polar molecules, as measured by the electronegativity difference $\Delta$EN, are expected to attract one another strongly and hence have small volume. As \figref{lightvsheavy}(b) shows, molecules with $\Delta$EN$> 1$ tend to have higher plasma frequency than those with  $\Delta$EN$ <1$. Combined with average $M_{\textrm{r}}$, we can further classify high-index materials broadly into three classes: relatively heavy and non-polar molecules (low $\omega_p$), light and polar molecules (high $\omega_p$), and intermediates (indeterminate $\omega_p$), which are less common than the other two. Based on such classification, we can understand and predict which materials have higher plasma frequency than others, as shown in \figref{lightvsheavy}(d) (the FOM here amounts to rearranging \eqref{metric} such that $FOM \equiv \left[ \frac{(n^2 -1)^2 \omega}{n n'} \right]^{1/3} \leq \omega_p^{2/3}$, to normalize out the dependence of frequency and dispersion on refractive index). Many of the materials with relatively high plasma frequencies are polar dielectrics that support phonon polaritons. They tend to be transparent in the visible or UV spectrum, but not further out in the IR. Conversely, materials with lower plasma frequencies are typically IR-transparent, but not around the visible spectrum. The inset of \figref{lightvsheavy}(d) further confirms this trend, showing that the optimal frequency (at which the FOM is maximized) generally lies in the UV or visible for materials with high plasma frequencies and IR for those with low plasma frequencies.

Apart from maximizing electron density, there are other considerations that can bring materials closer to the bound. As previously described, the ideal susceptibility has an infinitely sharp (delta function) $\Im \chi$ spectrum, which is ideally achieved by materials with a flat band structure. Intuitively, the dispersion in $\Im \chi$ mostly arises from from transitions from the valence to conduction band for semiconductors, which contributes to $\Im \chi$ at frequency corresponding to the difference in energy between the two bands for each crystal momentum [ref]. Thus, the more flat the band structure, the more sharply peaked is the corresponding $\Im \chi$ spectrum. Also, IR-inactive molecules are desirable at IR frequencies where phonon dispersion can become non-negligible. As depicted in \figref{lightvsheavy}(d), these factors determine to what extent each material comes close to the bound. For example, PbSe, InAs, and other IR-active molecules suffer from phonon dispersion and hence fall far short of the bound.
		    
Our analysis of the various factors described above provides useful insight into the ideal high-index material. First, the molecule should consist of relatively light elements, at least on average (small average $M_{\textrm{r}}$). Second, each molecule should consist of at least two or more elements with large $\Delta$EN. Third, the material should have a relatively flat band structure---ideally just two flat bands. Lastly, IR-inactive materials like silicon may be preferable for mid-IR frequencies, even if they do not satisfy the three preceding criteria.